\documentclass[conference]{IEEEtran}
\IEEEoverridecommandlockouts
\usepackage{cite}
\usepackage{amsmath,amssymb,amsfonts}
\usepackage{graphicx}
\usepackage{hyperref}
\usepackage{enumitem}
\usepackage{textcomp}
\usepackage{xcolor}
\def\BibTeX{{\rm B\kern-.05em{\sc i\kern-.025em b}\kern-.08em
    T\kern-.1667em\lower.7ex\hbox{E}\kern-.125emX}}
\usepackage{algorithm}
\usepackage[noend]{algorithmic}
\usepackage{svg}
\usepackage[outdir=./]{epstopdf}
\usepackage{listings}
\begin{document}
\newtheorem{thm}{Theorem}
\newtheorem{problem}[thm]{Problem}
\newcommand{\TRICOMMENT}[1]{\hfill$\triangleright$\textit{#1}}

\title{NeuroDB: A Neural Network Framework for Answering Range Aggregate Queries and Beyond}

\author{\IEEEauthorblockN{Sepanta Zeighami}
\IEEEauthorblockA{\textit{USC} \\
zeighami@usc.edu}
\and
\IEEEauthorblockN{Cyrus Shahabi}
\IEEEauthorblockA{\textit{USC} \\
shahabi@usc.edu}
}


\setlength{\textfloatsep}{0pt}
\setlength{\floatsep}{0pt}
\setlength{\intextsep}{0pt}
\setlength{\dbltextfloatsep}{0pt}
\setlength{\dblfloatsep}{0pt}
\setlength{\abovecaptionskip}{0pt}
\setlength{\belowcaptionskip}{0pt}

\maketitle

\begin{abstract}
Range aggregate queries (RAQs) are an integral part of many real-world applications, where, often, fast and approximate answers for the queries are desired. Recent work has studied answering RAQs using machine learning models, where a model of the data is learned to answer the queries. However, such modelling choices fail to utilize any query specific information. To capture such information, we observe that RAQs can be represented by \textit{query functions}, which are functions that take a \textit{query instance} (i.e., a specific RAQ) as an input and output its corresponding answer. Using this representation, we formulate the problem of learning to approximate the query function, and propose NeuroDB, a \textit{query specialized} neural network framework, that answers RAQs efficiently.
We experimentally show that NeuroDB answers RAQs orders of magnitude faster than the state-of-the-art on real-world, benchmark and synthetic datasets. Furthermore, 
NeuroDB is query-type agnostic (i.e., it does not make any assumption about the underlying query type) and our observation that queries can be represented by functions is not specific to RAQs. Thus, we investigate whether NeuroDB can be used for other query types, by applying it to distance to nearest neighbour queries. We experimentally show that NeuroDB outperforms the state-of-the-art for this query type, often by orders of magnitude. Moreover, the same neural network architecture as for RAQs is used, bringing to light the possibility of using a generic framework to answer any query type efficiently.  


\end{abstract}


\begin{sloppy}
\vspace{-0.1cm}
\section{Introduction}
\vspace{-0.1cm}
Range aggregate queries (RAQs) are an integral part of many real world applications. Calculating the total profit over a period from sales records or the average pollution level for different regions for city planing \cite{ma2019dbest} are examples of their use cases. Often, due to large volume of data, an exact answer takes too long to compute and a fast approximate answer is preferred. In such scenarios, there exists a time/space/accuracy trade-off, where algorithms can sacrifice accuracy for time or space. For example, consider a geospatial database containing latitude and longitude of location signals of individuals and, for each location signal, the duration the individual stayed in that location. A potential RAQ on this database is to calculate the average time spent by users in an area, which can be useful for analyzing different Points of Interest (POIs). Approximate answers within a few minutes of the exact answer can be acceptable in such applications. We use such a scenario as our running example, with the database shown in Fig. \ref{fig:visit_query_function} (left).

Research on RAQs has focused on improving the time/space/accuracy trade-offs.
Existing methods can be divided into sampling-based \cite{hellerstein1997online, agarwal2013blinkdb, chaudhuri2007optimized, park2018verdictdb} 
and model-based methods \cite{graham2012synopses, schmidt2002propolyne, ma2019dbest, thirumuruganathan2019approximate, hilprecht2019deepdb}. Sampling-based methods sample a subset of the database 
and answer the queries based on the samples. Model-based methods develop a model of the data to answer the queries. Models can be of the form of histograms, wavelets and data sketches (see \cite{graham2012synopses} for a survey). More recently, learning-based regression and density data models \cite{ma2019dbest, thirumuruganathan2019approximate, hilprecht2019deepdb} were proposed which have shown improvements over existing techniques. 

A major drawback of the recent learning-based approaches is that they learn a model from the data that is oblivious to the queries being asked. Learning from the data misses the opportunity to learn information about the queries that can help answer them faster. This information can be of two forms. (1) It can be about the query workload. In real-world databases, certain queries are far more common than others. In fact, OLAP systems (e.g., \cite{oracle, tableau}) divide attributes into \textit{measure} and \textit{dimension} categories where common RAQs have an aggregation function on a \textit{measure} attribute and range predicates on \textit{dimension} attributes. In our running example, visit duration can be defined as a measure attribute and lat. and lon. as dimension  attributes. Other queries, such as the query of average latitude given a visit duration range may not make any semantic sense. Even given the measure and dimension attributes, some ranges are queried more often than others. In our running example, more RAQs may be asked in downtown vs. a residential area. Diverting model capacity to learn about the queries that will actually be asked can improve the performance of a system. (2) The information can also be about the patterns in the answer to the query. The average visit duration for different places in the downtown of a city may be similar, while the answer will be different for suburban areas. A model can find such patterns, so that a compact representation of the data \textit{relevant to the queries} is learned.


\begin{figure}
    \centering
    \begin{minipage}{0.47\columnwidth}
    \includegraphics[width=\textwidth]{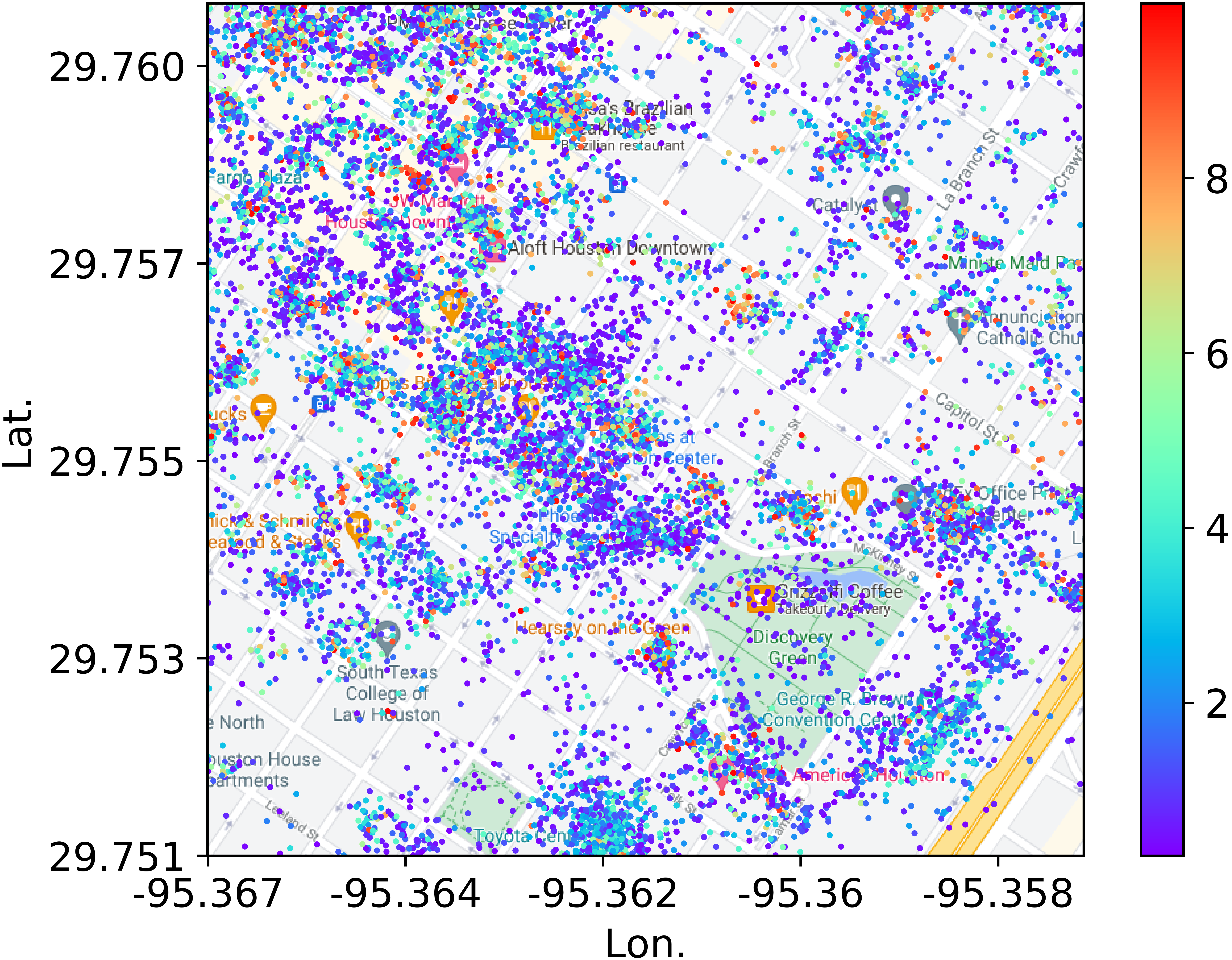}
    \end{minipage}
    \hfill
    \begin{minipage}{0.47\columnwidth}
    \includegraphics[width=\textwidth]{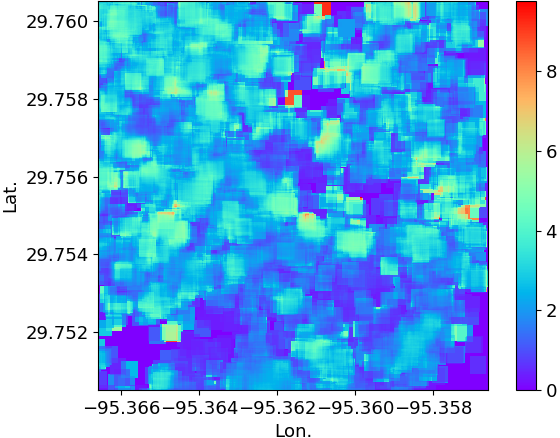}
    \end{minipage}
    \caption{Running example: (left) Database of location signals, and (right) the query function of average visit duration. Color shows visit duration in hours.}
    \label{fig:visit_query_function}
\end{figure}
\vspace{-0.1cm}
\subsection{NeuroDB for Range Aggregate Queries}\label{sec:intro:neurodb_RAQ}
\vspace{-0.1cm}
We propose a fundamentally different approach towards answering RAQs. First, we introduce the notion of a \textit{query function}. Intuitively, a range aggregate \textit{query function} takes as an input a range predicate and outputs a real number, which is an aggregation of a measure attribute. In our running example, consider the query of average visit duration for a 50m$\times$50m rectangle with bottom left corner at the geo-coordinate $(x, y)$. For such an RAQ, a query function, $f_D(x, y)$, can be defined that takes as input the geo-coordinate of the rectangle and outputs the average visit duration of data points in the rectangle. This query function is plotted in Fig.~\ref{fig:visit_query_function} (right). We call a particular input to the query function a \textit{query instance}. For example, Fig.~\ref{fig:visit_query_function} (right) shows that for query instance (-95.3615, 29.758) the answer is 9, i.e., $f_D(-95.3615, 29.758)=9$. 

The observation that RAQs can be represented as query functions allows us to learn models to answer them efficiently. To that end, we propose a novel framework, called \textit{Neural Databases} (NeuroDB), to efficiently answer RAQs. Specifically, let $q$ be a query instance issued by a user, and let $f_D(x)$ be the range aggregate query function, defining what the answer is for any query instance $x$ (and thus the ground truth answer for $q$ is $f_D(q)$). During a pre-processing step, NeuroDB uses solved query instances from a known algorithm to learn a model that approximates the query function well. 
The model takes a query instance as an input and outputs an answer, and the training objective is to optimize the model so that its answer is as close to ground truth as possible. At test time, NeuroDB outputs $\hat{f}_D(q;\theta)$ as its answer for query $q$, \textbf{eliminating both the database and the algorithm it learned from}. In our running example, NeuroDB learns a model that looks similar to Fig.~\ref{fig:visit_query_function} (right), but can be evaluated fast and without accessing the database at all. We use neural networks as our model and propose a neural network architecture that can be easily trained, answer the queries fast and accurately, and scale with dimensionality and data size.

NeuroDB improves the state-of-the-art in two ways. First, we experimentally observed that NeuroDB answers RAQs multiple orders of magnitude faster than state-of-the-art and with better accuracy. Thus, in traditional data management systems, NeuroDB can be an addition to the query processing engine and can considerably improve the performance of RAQs, while a default query processing engine can answer queries for which NeuroDB is not trained. In this regard, NeuroDB can be used similar to how indexes are used in database systems. Second, NeuroDB trained for a query function is typically much smaller than data size, while the model learned can be used to answer the queries without accessing the data. This is beneficial for applications requiring efficient release and storage of data. For instance, location data aggregators (e.g., SafeGraph \cite{safegraph}) can train NeuroDB to answer the average visit duration query, and release it to interested parties instead of the dataset. This can improve the storage, transmission and query processing costs for all parties. 



\subsection{Beyond Range Aggregate Queries}\label{sec:intro_NN}
Our observation that queries can be represented by query functions is not specific to RAQs, and is true for any query type. Considering such representation opens the path for using learned models to answer different database query types. Hence, an important question is to understand the utility of learned models for different query types. We take the first step in this direction by applying NeuroDB to another query type. Specifically, we consider two variants of $k$-th nearest neighbour queries. This is possible because the training of NeuroDB is query type agnostic, that is, it does not make assumptions about the query type during training. 

First, we consider $k$-th nearest neighbour queries ($k$-NN), which, for a $d$-dimensional database $D$, can be represented by a function, $f_D^{NN}(.)$ that takes a $d$-dimensional point as an input, and outputs another $d$-dimensional point. NeuroDB can be used to train a model, $\hat{f}_D^{NN}(.;\theta)$ that mimics the function $f_D^{NN}(.)$. We observed that NeuroDB can provide an answer for the $k$-NN query that is spatially close to the true answer. However, for a query $q$, the output of $\hat{f}_D^{NN}(q;\theta)$ is not necessarily in the database (but rather some point in $\mathcal{R}^d$ whose distance to $f_D^{NN}(q)$ is small), while the answer to nearest neighbour query needs to be a point in the database. 
Such an answer may not have much utility in real-world applications. 


Next, we consider the \textit{distance to} $k$-th nearest neighbour query, where the query answer is the distance to and not the actual point in the database. This query can be represented by a function that takes a $d$-dimensional query point as an input and outputs a real number. Distance to nearest neighbour query is useful for various applications such as active learning \cite{settles2009active, fujii1999selective, xu2007incorporating} (e.g., as a diversity score for selecting samples for training a model), outlier detection \cite{chandola2009anomaly} and assessing probability of getting infected from a disease (e.g., COVID-19) calculated based on proximity to the nearest infected person. We observed that NeuroDB is able to answer distance to $k$-NN queries orders of magnitude faster than state-of-the-art, while only taking space that is a small fraction of data size. NeuroDB achieves this because 
it answers distance to $k$-NN queries without calculating any distances or finding any of the neighbours at query time. 
Instead, NeuroDB approximately materializes the query results, with higher probability regions (according to query distribution) being materialized more accurately. This allows the framework to perform especially well for high-dimensional data where queries are from a small portion of the total high dimensional space (e.g. images\cite{roweis2000nonlinear}).
%
%

The results in this paper suggest dividing database queries into \textit{item queries} and \textit{statistic queries} categories. Item queries are queries where the answer is an actual database item, while statistic queries are queries where the answer is a numerical statistic calculated based on the database. We conjecture that NeuroDB, on its own, is useful for statistic queries. Studying this conjecture for different statistic queries is an interesting future research direction, where a generic framework can be used to answer multiple different query types in this category. This paper is a first step in that direction, where a framework designed for range aggregate queries is shown to also efficiently answer distance to the $k$-th nearest neighbour queries and outperform the state-of-the-art for both of them. Although NeuroDB outperforming the state-of the-art in either of the two query types is significant on its own, this paper also sheds light on the possibility that a learned framework can be generalizable to multiple query types. As such, it can improve the performance of a system while also saving time when designing it. For item queries, NeuroDB may be useful as part of other algorithms designed for such queries, the study of which we leave for future work.

\subsection{Contributions and Roadmap}
\begin{itemize}[leftmargin=10pt]
        \item We formulate the problem of learning RAQs with function approximators (Sec. \ref{sec:def}); 
        \item Propose NeuroDB, the first neural network framework to answer RAQs efficiently  (Secs. \ref{sec:NeuroDB}, \ref{sec:raq_extension}); and
        \item Show how NeuroDB can also answer distance to nearest neighbour queries (Sec. \ref{sec:NN}).
        \item Our experiments show that
        \begin{enumerate}[leftmargin=4pt]
            \item {NeuroDB} enjoys orders of magnitude gain in query time and provide better accuracy over state-of-the-art for answering RAQs using real-world, TPC-benchmark and synthetic datasets (Sec.~\ref{sec:exp:rang_agg}); and
            \item The same architecture is used to answer distance to nearest neighbour query with orders of magnitude gain in query time over state-of-the-art on real datasets. NeuroDB's query time is not affected by $k$ and only marginally impacted by data dimensionality (Sec. \ref{sec:exp:dist_NN}).
        \end{enumerate}
        \end{itemize}
\noindent Sec.~\ref{sec:related_work} presents related work and we conclude in Sec.~\ref{sec:conclusion}.

\section{Problem Definition}\label{sec:def}
\noindent\textbf{Range Aggregate Queries}. Consider a dataset (or table) $D$ with $n$ records and with $d$ attributes, $X_1$, ..., $X_d$. For ease of discussion, we start by considering range aggregate queries that can be represented by the following SQL query. Such a query captures many real-world RAQs \cite{ma2019dbest}. We discuss the extension of our solution to more general RAQs in Sec.~\ref{sec:raq_extension}. 

\begin{tabular}{c}
\centering
\hspace{-0.7cm}
\begin{lstlisting}
SELECT $AGG$($X_m$) FROM $D$
WHERE ($x_1^l \leq X_1  < x_1^u $) AND ... AND ($x_d^l \leq X_d  < x_d^u $)
\end{lstlisting}
\end{tabular}

In the above SQL statement, for any $i$, $x_i^l$ and $x_i^u$ are \textit{query variables}, and they, respectively, define lower and upper bounds on the range of the attributes. For any $i$, $x_i^l$ and $x_i^u$ can be $-\infty$ and $\infty$ respectively, in which case there are no restriction on the values of $X_i$ in the query. We say that an attribute is \textit{not active} in the query in that case, and is \textit{activate} otherwise. Furthermore, $AGG$ is a user defined \textit{aggregation function} (typical examples include $SUM$, $AVG$ and $COUNT$), and $X_m$, for $1\leq m\leq d$, is the \textit{measure attribute}. For ease of discussion, we assume the measure attribute and the aggregation function are fixed, that is, we are only interested in answering RAQs with measure attribute $X_m$ and aggregation function $AGG$. We relax this assumption when discussing general RAQs in Sec.~\ref{sec:raq_extension}. Furthermore, let $q=(x_1^l, ..., x_d^l, x_1^u, ..., x_d^u)$ be a $d_{pred}$-dimensional vector for $d_{pred}=2\times d$. We call $q$ a \textit{query instance} and we assume query instances follows some distribution $\mathcal{Q}$. Thus, different query instances correspond to different range predicates for the measure attribute $X_m$ and aggregation function $AGG$.

We use a real-world database of location signals as our running example. The database, shown in Fig.~\ref{fig:visit_query_function} (left) contains latitude and longitude of GPS signals obtained from cell-phone devices and, for each location signal, the duration the user stayed in that location (more details about the dataset are provided in Sec. \ref{sec:exp}). The dataset is for downtown Houston, plotted on the map obtained from \url{maps.google.com}. Setting $X_1$, $X_2$ and $X_3$ to represent lat., lon. and visit duration attributes, respectively and $m=3$ and $AGG$ as $AVG$, the SQL statement above is the query of average visit duration for check-ins that fall in a rectangle with bottom left corner at $(x_1^l, x_2^l)$ and top right corner at $(x_1^u, x_2^u)$, and whose visit duration is between $x_3^u$ and $x_3^l$. The distribution $\mathcal{Q}$ decides which query instance are more common than others. $\mathcal{Q}$ may ensure $x_3^l=-\infty$ and $x_3^u=\infty$ with probability 1, since visit duration is typically not an active attribute, or, $(x_1^l, x_2^l)$ and $(x_1^u, x_2^u)$ may follow a distribution so that the range is more often around POIs in downtown rather than residential areas.



\noindent\textbf{Query Functions}. Such RAQs can be represented by a function of the query instance. 
Define the function $f_D(.)$ so that for a query $q$, $f_D(q)$ is the answer to the above SQL statement. We call $f_D(.)$ a \textit{query function}. A \textit{query instance} defines a particular query on a database, and a \textit{query function} is a function that maps a query instance to its answer. In our running example, the function $f_D(.)$ takes as input the ranges on three attributes (and thus, its input dimensionality is 6) and outputs a real number. In Fig. \ref{fig:visit_query_function} (right), we show this function for a subset of query instances. Specifically, it shows the query of average visit duration given a square of side length 0.00043 in geo-coordinates (which is about 50m) with bottom left corner at location $(x_1^l, x_2^l)$, which is achieved by setting $x_3^l=-\infty$, $x_3^u=\infty$, $x_1^u=x_1^l+0.00043$, $x_2^u=x_2^l+0.00043$.



\noindent\textbf{Problem Statement}. Our goal in this paper is to learn a function approximator, $\hat{f}_D(.; \theta)$ that approximates the query function, $f_D(.)$, well. In the general problem setting, $\hat{f}_D(q; \theta)$ can be any algorithm, from a combinatorial methods that operates on the data to a neural network. For any such function approximator, let $\Sigma(\hat{f}_D)$ be its storage cost (e.g., for neural networks, number of parameters) and $\tau(\hat{f}_D)$ be its evaluation time or query time (e.g., for neural networks, the time it takes for a forward pass). Furthermore, let $\Delta(f(q), y)\geq 0$ be an error function that measures how bad a solution $y$ is when the actual answer is $f(q)$ for a query $q$, e.g., it can be defined as 0-1 loss or $\lVert f_D(q)-y\rVert$. The problem studied in this paper is learning to answer range aggregate queries with time and space constraints, formulated as follows.

\begin{problem}\label{prob:RAQ}
Given a query function $f_D(.)$, query distribution $\mathcal{Q}$, class of function approximators, $\Theta$, and time and space requirements $t$ and $s$, find 
$$\hspace{-0.1cm}\arg\min_{\theta\in \Theta} E_{q\sim \mathcal{Q}}[\Delta(f_D(q), \hat{f}_D(q;\theta))]  \text{ s.t. }\Sigma(\hat{f}_D)\leq s \text{, } \tau(\hat{f}_D)\leq t$$
\end{problem}

In the problem formulation, $\Theta$ shows our modelling choice. 
Furthermore, since we usually only have access to a set $Q$ of samples from $\mathcal{Q}$ but not the distribution, we aim at optimizing $\frac{1}{|Q|}\sum_{q\in Q}\Delta(f_D(q), \hat{f}_D(q;\theta)$ instead of $E_{q\sim \mathcal{Q}}[\Delta(f_D(q), \hat{f}_D(q;\theta)]$.

\begin{figure*}
    \centering
    \includegraphics[width=\textwidth]{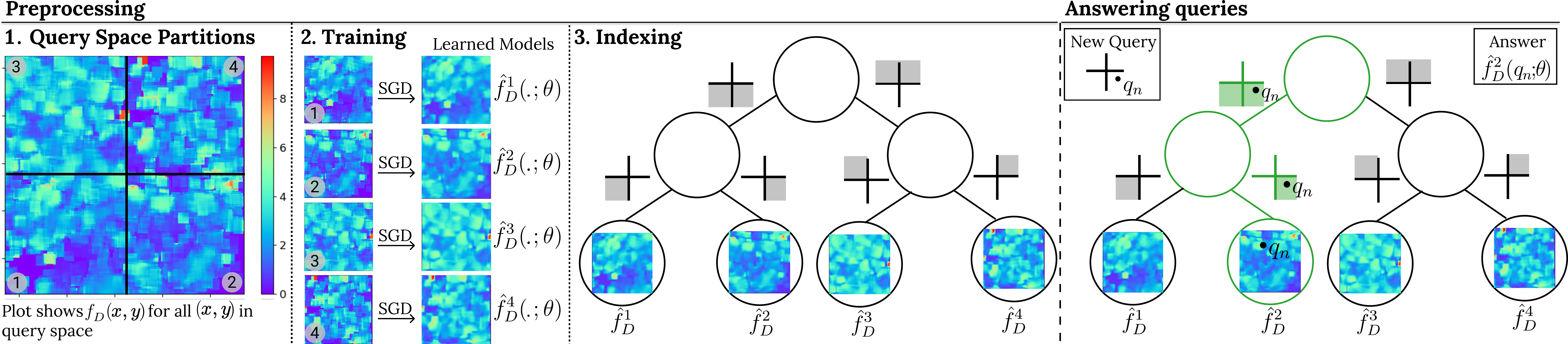}
    \vspace{-0.5cm}
    \caption{NeuroDB Framework}
    \label{fig:ndb}
\end{figure*}
\section{NeuroDB}\label{sec:NeuroDB}
To solve Problem \ref{prob:RAQ}, we design and train a function approximator in Sec.~\ref{sec:arch}. We show how it can answer queries in Sec.~\ref{sec:answering_queries} and analyze its performance in Sec.~\ref{sec:analysis}.


\subsection{Model Architecture and Training}\label{sec:arch}
\subsubsection{Challenges}\label{sec:arch:challanges} Using neural networks to answer RAQs comes with its own challenges. We first discuss these challenges that motivate our design choices. 

\noindent(1) \textbf{Query time/accuracy trade-offs}. To improve the accuracy of a neural network, we need to increase the number of its parameters. Meanwhile, a forward pass of a neural network takes time linear in the number of its parameters (assuming no parallelization). A design is needed that can limit the increase in query time while improving accuracy. We avoid parallelization for fair comparison with existing methods, but its implications are discussed in Appendix~\ref{appx:parallel}.



\noindent(2) \textbf{Dependence on $n$}. The query function $f_D$ is dependant on the dataset $D$ and the complexity of approximating $f_D$ depends on the dataset and its size $n$. $f_D$ is a piece-wise constant function and increasing $n$ can increase the number of pieces. Intuitively, this is because the answer to a query changes only when the set of matching points to the range predicate changes. Thus, points of discontinuity can happen whenever a data point is on the boundary of the range predicate, as in those scenarios changing the range predicate by any non-zero amount can change the value of the query function by a fixed amount. This also implies that the number of discontinuities is larger when there are more points in a database. 
Thus, our proposed architecture should be able to take data size into account. Fig.~\ref{fig:visit_query_function} (right) shows the piece-wise constant nature of the query function in our running example.

\noindent(3) \textbf{Training time}. Since number of parameters needs to increase with accuracy and data size, training could become harder and take more time and space. Although training time is a preprocessing step and does not affect query time, training of the network should be feasible with existing GPUs.

\subsubsection{Architecture and Training} 
NeuroDB consists of multiple neural networks with identical structures. At a preprocessing step, we split the query space into a number of partitions and learn a different neural network to answer the queries for each partition. At query time, to answer a new query, we find which neural network is responsible for it and then perform a forward pass of that neural network. The answer to the query is the output of the neural network.

Fig.~\ref{fig:ndb} shows this framework for our average visit distribution RAQ example. Here, based on the formulation in Sec.~\ref{sec:def}, the query space is the possible bottom-left co-ordinates of the query rectangle. Fig.~\ref{fig:ndb} shows that in a preprocessing step, the query space is partitioned into 4 partitions. Then, for the $i$-th partition, a model, $\hat{f}_D^i(.; \theta)$, is learned that mimics the query function $f_D(.)$ for that part of the query space. Subsequently, the partitions are indexed. Finally, given a new query, $q_n$, the index is traversed to find which partition it belongs to. A forward pass of the corresponding neural network is performed and the output is returned as the answer, in this case  $\hat{f}_D^2(q_n; \theta)$.





Before explaining the specifics of our design, we discuss how this architecture addresses the challenges of Sec. \ref{sec:arch:challanges}. (1) Query time depends on the time it takes to find the partition the query belongs to, $t_p$, and the time of a forward pass for the neural networks, $t_n$. Using our architecture, increasing number of partitions increases $t_p$ but not $t_n$. Given that indexing can be used to search the partitions, $t_p$ is generally very small and increasing it has negligible impact on query time. As a results, we can increase number of parameters and model capacity at a low cost for query time. (2) Number of partitions can be used to increase the size of the architecture with data size and can be set as a function of $n$. (3) Training can be done independently for each neural network used. The benefits of this is twofold. First, neural networks can be trained in parallel and even on different devices, which speeds up training. Second, training requires less memory because all the networks do not need to be loaded at once. Thus, we can train only as many networks as there is memory for, as opposed to having to train all the network at once which requires larger memory.

To summarize, this approach requires a method for partitioning the space and indexing them, as well as designing a neural network and training it for each partition. Our method uses a kd-tree to partition the space and index them,  and these steps are performed together. Thus, we first discuss those two steps and then discuss training of the neural networks.

\noindent\textbf{Partitioning}. For this method to be successful, a good partitioning method needs to be chosen. Although it may be possible to learn the partitioning, our experiments showed that learning the partitioning is difficult and computationally intensive in practice. 
Instead, we take a different approach towards partitioning the space. Recall that our objective is to minimize $E_{q\sim \mathcal{Q}}[\Delta_{f_D}(q, \hat{f}_D(q;\theta)]$ which is $\sum_i p(q\in P_i)E_{q\sim \mathcal{Q}_i}[\Delta_{f_D}(q, \hat{f}_D(q;\theta)]$, where $P_i$ is the $i$-th partition and $\mathcal{Q}_i$ is the distribution of queries in partition $P_i$ (i.e., if $g_\mathcal{Q}(q)$ is p.d.f. of $\mathcal{Q}$, p.d.f. of $\mathcal{Q}_i$ is $\frac{g_\mathcal{Q}(q)}{p(q\in P_i)}$ if $q\in P_i$ and 0 otherwise). Thus, for each partition, its contribution to our objective is dependant on probability of it being queried as well as the average approximation error, where the former depends on the query distribution while the latter depends on the complexity of the function being approximated. Hence, we should select partitions such that high probability areas are approximated accurately, while the error for low probability partitions can be higher. We use the general observation that reducing the size of the space approximated by a neural network allows for better approximations in the smaller space. We choose partitions that are smaller where the queries are more frequent and larger where they are less frequent. This can be done by partitioning the space such that all partitions are equally probable. 

To this end, we build a kd-tree on our sampled query set, $Q$. Note that the split points in the kd-tree can be considered as estimates of the median of the distribution $\mathcal{Q}$ (conditioned on the current path from the root) along one of its dimensions obtained from the samples in $Q$. We build the kd-tree by specifying a maximum height, $h$, and splitting every node until all leaf nodes have height $h$, which creates $2^h$ partitions. Splitting of a node $N$ is done based on median of one of the dimensions of the subset, $N.Q$, of the queries, $Q$, that fall in $N$. Thus, a leaf node will have at least $\lfloor\frac{|Q|}{2^{h-1}}\rfloor$ queries. The complete procedure is shown in Alg.~\ref{alg:get_index}. To build an index with height $h$ rooted at a node, $N_{root}$ (note that $N_{root}.Q=Q$), we call $get\_index(N_{root}, h, 0)$ defined in Alg.~\ref{alg:get_index}. 


\begin{algorithm}[t]
\begin{algorithmic}[1]
\REQUIRE A kd-tree node $N$, tree height $h$ and dimension, $i$ to split the node, $N$ on 
\ENSURE A kd-tree with height $h$ rooted at $N$
\IF{$h=0$} 
    \STATE $N.model \leftarrow train\_model(N.Q, f_D(N.Q))$\label{alg:get_index:train}
    \RETURN 
\ENDIF
\STATE $N.val\leftarrow$ median of $N.Q$ along $i$-th dimension
\STATE $N.dim\leftarrow i$
\STATE $Q_{left}=\{q|q\in N.Q, q[N.dim]\leq N.val\}$
\STATE $Q_{right}=\{q|q\in N.Q, q[N.dim]>N.val\}$
\FOR{$x\in \{left, right\}$}
    \STATE $N_x$ $\leftarrow$ new node
    \STATE $N_x.Q\leftarrow Q_x$
    \STATE $N.x\leftarrow N_x$\TRICOMMENT{Adding $N_x$ as left or right child of $N$}
    \STATE $get\_index(N_x, h-1, (N.dim+1)\mod d_{pred})$
\ENDFOR
\end{algorithmic}
\caption{$get\_index(N, h, i)$}\label{alg:get_index}
\end{algorithm}

\noindent\textbf{Training}. We train an independent model for each of the $2^h$ leaf nodes. For a leaf node, $N$, this happens in Line \ref{alg:get_index:train} of Alg.~\ref{alg:get_index}, where the function $train\_model(N.Q, f_D(N.Q))$ returns a trained model on the training data $N.Q$ and $f_D(N.Q)$. $N.Q$ are query samples that fall within the part of the query space that node $N$ is responsible for. $f_D(N.Q)$ is used as the training label for queries in $N.Q$. Note that the answers can be collected through any known algorithm, where a typical algorithm iterates over the points in the database, pruned by an index, and for a candidate data point checks whether it matches the RAQ predicate or not.
We emphasize that this is a pre-processing step. That is, this sample collection step is only performed once and is only used to train our model. Furthermore, the process is embarrassingly parallelizable across training queries, if preprocessing time is a concern. 


The process of training is similar to a typical supervised training of a neural network with stochastic gradient descent (SGD). We use Adam optimizer \cite{kingma2014adam} for training and use a squared error loss function, that is, for the $i$-th partition corresponding to the leaf node $N$, the minimization objective is $\frac{1}{|N.Q|}\sum_{q\in N.Q}(f_D(q)-\hat{f}_D^i(q;\theta))^2$. 
\begin{algorithm}[t]
\begin{algorithmic}[1]
\REQUIRE kd-tree root node $N$ and query $q$
\ENSURE Answer to $q$
\WHILE{$N$ is not leaf}
    \IF{$q[N.dim] \leq N.val$}
        \STATE $N \leftarrow N.left$
    \ELSE
        \STATE $N \leftarrow N.right$
    \ENDIF
\ENDWHILE
\RETURN $N.model.forward\_pass(q)$
\end{algorithmic}
\caption{$get\_answer(N, q)$}\label{alg:get_answer}
\end{algorithm}

\noindent\textbf{Neural Network Architecture}. We use a fully connected neural network for each of the partitions. The architecture is the same for all the partitions and consists of $n_l$ layers, where the input layer has dimensionality $d_{pred}$, the first layer consists of $l_{first}$ units, the next layers have $l_{rest}$ units and the last layer has 1 unit. We use relu activation all the layers (except the output layer). 
Note that $n_l$, $l_{first}$ and $l_{rest}$ are hyper-parameters of our model. Although approaches in neural architecture \cite{zoph2016neural} search can be applied to find them, they are generally computationally expensive. In this paper, we use a simple heuristic. We select one of the partitions, and do a grid search on the hyper-parameters. Since our neural network architecture for each partition is small, this grid search can be done in a practical time frame. The grid search finds the hyper-parameters so that NeuroDB satisfies the space and time constraints in Problem~\ref{prob:RAQ} while maximizing its accuracy. 



\subsection{Answering Queries}\label{sec:answering_queries}
Answering queries using our NeuroDB framework is simple. The pseudocode is shown in Alg.~\ref{alg:get_answer}. For a query, $q$, first, the kd-tree is traversed to find the leaf node that the query $q$ falls into. The answer to the query is a forward pass of the neural network corresponding to the leaf node.

\vspace{-0.2cm}
\subsection{Analysis}\label{sec:analysis}
Given a value of $h$, there are $2^h$ partitions and each contains a neural network. We let $h=\log \frac{n}{c}$ for some user parameter $c$ denoting the \textit{capacity} of a neural network, so that the number of partitions, $N_p$, is $\frac{n}{c}$. That is, we increase the number of partitions linearly in $n$. 
Intuitively, the capacity of a neural network denotes how \textit{complex} of a function it can approximate well, which depends on the number of neural network parameters, number of points in the database (as well as their distribution) and number of training samples available. We leave a theoretical study of the capacity of a neural network to the future work, but briefly mention that recent work on \textit{memorization capacity} of a neural network \cite{yun2019small} can be seen as a first step in this direction. We revisit the topic of what value of $h$ should be chosen empirically in our experiments. Regarding time and space complexity, for a fixed neural network architecture but variable data size and dimensionality, there will be $O(d_{pred}n)$ number of parameters, which means the space complexity is $O(d_{pred}n)$. Furthermore, at query time, the kd-tree needs to be traversed which takes $O(\log n)$, and a forward pass of the neural network takes $O(d_{pred})$. Thus, time complexity of the algorithm is $O(\log n +d_{pred})$. 

We acknowledge that, similar to recent learning-based approaches \cite{hilprecht2019deepdb, ma2019dbest} we do not provide an analytical formulation of the accuracy of NeuroDB. Nevertheless, we experimentally show that this architecture can provide good accuracy in practice. In our experiments, we discuss how model accuracy depends on number of training samples available, number of partitions and the size of each neural network.  

Regarding training time, given a network architecture, and assuming $T$ iterations of stochastic gradient descent, training time can be quantified as $O(d_{pred}nT)$, where each iteration for each network takes $O(d_{pred})$ and there are $O(n)$ networks.

\section{NeuroDB for General RAQs}\label{sec:raq_extension}
To train NeuroDB, we require a query function $f_D(.)$ and a query set $Q$. NeuroDB treats the query function as a black box, only utilizing it to collect the labels for training samples in $Q$. Thus, we call NeuroDB a \textit{query-type agnostic} framework. Subsequently, after NeuroDB is trained for a query function, it can be used to answer corresponding query instances. As such, to extend NeuroDB to other RAQs, we only need to show how they can be represented as a query function. After such a query function representation is created, we can learn NeuroDB to answer the corresponding query instances. Below, we first discuss how this representation can be obtained in general RAQ settings, then we discuss how NeuroDB can be applied to answer RAQs in real-world databases. 



\subsection{Query Representation}
\noindent\textbf{General RAQs}. First, we generalize our definition of RAQs. An RAQ consists of a range predicate, and an aggregation function $AGG$. We consider range predicates that can be represented by a \textit{query instance} $q$, a $d_{pred}$ dimensional vector, and a binary \textit{predicate function}, $P_f(q, x)$, that takes as inputs a point in the database, $x$, $x\in D$, and the query instance $q$, and outputs whether $x$ matches the predicate or not. Furthermore, $AGG$ is a function that takes the set of matching data points to the query as an input and outputs a real number. Note that the notion of a measure attribute is implicitly captured in the definition of aggregation function (e.g., $AVG(X_m)$ can be defined by the aggregation function $AGG(S) = \sum_{s\in S}\frac{s[X_m]}{|S|}$, where $s[X_m]$ is the value of $X_m$ attribute for a record $s$). Then, given a predicate function and an aggregation function, range aggregate queries can be represented by the query function $f_D(q)=AGG(\{x|x \in D, p(x, q)=1\})$.

The above formulation divides the set of all possible RAQs into a set of different query functions, each defined for a specific aggregation and predicate function. In a real-world application, relevant query functions can be created from the query workload, where RAQs with the same predicate and aggregation function but different query instances can be used to define a query function. We avoid specifying how the predicate function should be defined to keep our discussion generic to arbitrary predicate functions, but some examples follow. To represent the RAQs of the form discussed in Sec.~\ref{sec:def}, $q$ can be defined as lower and upper bounds on the attributes and $P_f(q, x)$ defined as the WHERE clause in Sec.~\ref{sec:def}. We can also have $P_f(q, x) = x[1]>x[0]\times q[0]+q[1]$, so that $P_f(q, x)$ and $q$ define a half-space above a line specified by $q$. Furthermore, for many applications, WHERE clauses in SQL queries are written in a parametric form \cite{sql_microsoft, sql_nodepostgres, sql_dataworld}  (e.g., WHERE $X_1> ?param1$ OR $X_2> ?param2$, where $?param$ is the common SQL syntax for parameters in a query). Such SQL queries can readily be represented as query functions by setting $q$ to be the parameters of the WHERE clause. 




\noindent\textbf{Join and Group By Clauses}. Finally, although answering arbitrary SQL queries is not the focus of this papers, we discuss how RAQs with Join and Group By clauses can also be represented as query functions. If an SQL statement contains Join of tables, we can consider the dataset, $D$, on which the query function is defined to be the Joined tables. Then, the query function can be defined the same way as before, but operating on the joined tables. Furthermore, answering RAQs with a Group By clause using NeuroDB can be done similar to \cite{ma2019dbest}. Consider an RAQ with predicate function $P_f(q, x)$ and with the clause Group By G, where G is an attribute that takes one of $g_1$, ..., $g_k$ values. Such a query can be treated as $k$ different query functions (for each a different model can be learn) where the range predicate for the $i$-th query function, $P_f^i(q, x)$, is defined as $P_f(q, x)\land x[G]==g_i$. 

\subsection{Applying NeuroDB}\label{sec:applying_neurodb}
In our approach, possible RAQs on a database are divided into various query function. Subsequently, we learn different models for different query functions. We call this \textit{query specialization}, where specialized models are trained to answer specific query functions. This is beneficial because, as shown in our experiments, \textit{small specialized models} can answer a query function within microseconds, when the state-of-the-art \textit{non-specialized models} take milliseconds. Intuitively, this huge advantage is due to a specialized model being able to divert all its capacity to learn patterns for a specific query function. 

When a system requires answers to  multiple query functions, the choice of which query functions to learn a model for can be done by the database administrator, akin to the choice of which attribute to index. Intuitively, models should be learned for queries that are frequently performed on the database. Similar to existing work \cite{park2018verdictdb, ma2019dbest}, if a model has not been learned for a particular query function, we assume that a default query processing engine can answer the query. Overall, Learning a model for a query function increases space consumption (for model storage) but improves query time (since learned models can answer a query faster). Given that, in our approach, models take space equal to only a fraction of data size but speed up query time by orders of magnitude, learning multiple models to efficiently answer different query functions proves to be a realistic and beneficial choice. 

Furthermore, in contrast to recent learning-based methods \cite{hilprecht2019deepdb, ma2019dbest}, our approach can learn to answer different RAQs without making assumptions about the query function (besides the fact that the range predicate can be represented by a predicate function). This allows learning arbitrary and potentially application specific RAQs that generic database systems aren't optimized for (e.g., an arbitrary aggregation function on a polygon-shaped range predicate), and can substantially improve the performance of such systems. 

\begin{figure}
    \centering
    \begin{minipage}{0.22\textwidth}
    \includegraphics[width=\textwidth]{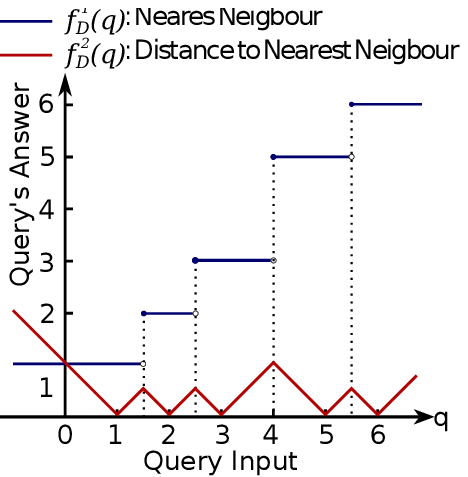}
    \end{minipage}
    \hfill
    \begin{minipage}{0.22\textwidth}
    \includegraphics[width=\textwidth]{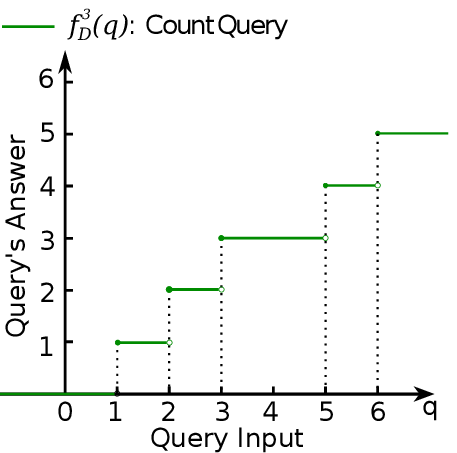}
    \end{minipage}
    \caption{Nearest neighbour, $f_D^1(q)$, distance to nearest neighbour, $f_D^2(q)$, (left) and range aggregate, $f_D^3(q)$, (right) query functions}
    \label{fig:query_functions}
\end{figure}

\section{NeuroDB for Nearest Neighbour Queries}\label{sec:NN}
In this section, we show how the NeuroDB framework can applied to nearest neighbour queries. As discussed in Sec.~\ref{sec:raq_extension}, NeuroDB framework is query type agnostic, and to apply NeuroDB to a query type, it is only required that the query type is represented by query functions. Thus, below we discuss the query representation for two variants of nearest neighbour queries. After specifying the query representation, NeuroDB can readily be applied to answer the queries. 
\noindent\textbf{$k$-th nearest neighbour query}. Given a $d$-dimensional database, an integer $k$, and a query point $q\in \mathcal{R}^d$, $k$-th nearest neighbour query is to find the point in $D$ whose distance to $q$ is the $k$-th smallest. For this query, we define $f_D(q):\mathcal{R}^d\rightarrow\mathcal{R}^{d}$ as the query function that takes in a point in $\mathcal{R}^d$ and outputs another point in $\mathcal{R}^d\cap D$. 

For the one dimensional database $D=\{1, 2, 3, 5, 6\}$, Fig.~\ref{fig:query_functions} (left) shows the $k$-th nearest neighbour query function for $k=1$ (Shown as $f_D^1(q)$ in the left figure). Observe that $f_D^1(q)$ is a step function, where the output of $f_D^1(q)$ is always a point in the database. $f_D^1(q)$ is constant over the query inputs whose nearest neighbour is the same, and points of discontinuity occur when the nearest neighbour changes (this happens at the mid-point between consecutive data points, shown by dashed lines in the figure). We have also included Fig. \ref{fig:query_functions} (right) for comparison with an RAQ. In Fig.~\ref{fig:query_functions} (right), we consider the count aggregation function and the range predicate is defined by the query variable $q$, $q\in R$, and predicate function $P_f(q, x)$, where $P_f(q, x)$ is set to 1 if $x>q$ and 0 otherwise (i.e., the query input specifies the beginning of the range and all ranges are of the form $(q, \infty)$).



\noindent\textbf{Distance to $k$-th nearest neighbour}. In addition to $k$-th nearest neighbour query, we also consider the \textit{distance to the $k$-th nearest neighbour} query. That is, if $p$ is the $k$-th nearest neighbour of a query point $q$, then $f_D(q)=d(p,q)$ for some distance metric, $d(p, q)$. Note that $f_D:\mathcal{R}^d\rightarrow\mathcal{R}$. 
As discussed in Sec.~\ref{sec:intro_NN}, this query function is useful for various application, but has not been traditionally studied separately from the nearest neighbour query, because combinatorial methods use a nearest neighbour algorithms to find the distance to the nearest neighbour. However, using neural networks we can directly calculate distance to the nearest neighbour without ever finding any of the nearest neighbours. 
The red curve in Fig. \ref{fig:query_functions} shows this query for $d=1$ and $k=1$ (on the same database mentioned above). Observe that, in contrast to the nearest neighbour query, this query is continuous. Furthermore, its range is a single real number, as opposed to $\mathcal{R}^{d}$. 

For both $k$-NN and distance to $k$-NN query types, following our framework of query specialization discussed in Sec.~\ref{sec:applying_neurodb}, we define $f_D(q)$ to be the query function for a specific value of $k$. Thus, $k$-NN query type is divided into different query functions, each for a value of $k$. The same discussion as in Sec.~\ref{sec:applying_neurodb} for RAQs applies to $k$-NN and distance to $k$-NN queries, e.g., we can learn NeuroDB for common values of $k$ and a default nearest neighbour algorithm can be used to answer the queries for values of $k$ where no model is learned. 

\section{Empirical Study}\label{sec:exp}
In this section, we first empirically evaluate NeuroDB for RAQs in Sec.~\ref{sec:exp:rang_agg}. Then, we apply the same NeuroDB architecture to answer distance to nearest neighbour queries in Sec.~\ref{sec:exp:dist_NN}.

\noindent\textbf{System Setup.} The experiments are performed on a machine running Ubuntu 18.04 LTS equipped with an Intel i9-9980XE CPU (3GHz), 128GB RAM and a GeForce TRX 2080 Ti NVIDIA GPU.

\noindent\textbf{Model Training and Evaluation}. For all the experiments, building and training of NeuroDB is performed in Python 3.7 and Tensorflow 2.1. Training of the model is done on GPU. The model is saved after training. For evaluation, a separate program written in C++ and running on CPU loads the saved model, and for each query performs a forward pass on the model. Thus, model evaluation is done with C++ and on CPU, without any parallelism for any of the algorithms. We refer to our algorithm as {NeuroDB}. Unless otherwise stated, we set the model depth to 5 layers, with the first layer consisting of 60 units and the rest with 30 units. The height of the kd-tree is set to 4. We found this architecture to provide both good accuracy and query time, but we also present results on impact of each of the hyperparamters.

\subsection{Range Aggregate Query}\label{sec:exp:rang_agg}
\subsubsection{Setup}\hfill\\
\noindent\textbf{Dataset}. We use real, benchmark and synthetic datasets for evaluation. The data size and dimensionality for each dataset is shown in Table~\ref{tab:dataset_size}. For each dataset, we specify a measure attribute on which the aggregation function operates.


\noindent\textit{PM2.5}. PM2.5 \cite{liang2015assessing} contains PM2.5 statistics for locations in Beijing. Similar to \cite{ma2019dbest}, we let PM2.5 to be the measure attribute. 

\noindent\textit{TPC-DS}. We used TPC-DS \cite{nambiar2016the} with scale factors 1 and 10, respectively referred to as TPC1 and TPC10. Since we study range aggregate queries, we consider the numerical attributes in store\_sales table as our dataset. We use net\_profit as the measure attribute. 

\noindent\textit{Synthetic datasets}. To study the impact of data dimensionality and distribution, we generated synthetic 5, 10 and 20 dimensional data from Gaussian mixture models (GMM) with 100 components whose mean and co-variance are selected uniformly at random, respectively referred to as G5, G10 and G20. GMMs are often used to model real data distributions \cite{reynolds2009gaussian}. We set one of the columns to be the measure attribute. 

\noindent\textit{Veraset}. As was used in our running example, we use Veraset dataset, which contains anonymized location signals of cell-phones across the US collected by Veraset \cite{veraset}, a data-as-a-service company. Each location signal contains an anonymized id, timestamp and the latitude and longitude of the location. We performed stay point detection \cite{ye2009mining} on this dataset (to, e.g., remove location signals when a person is driving), and extracted location visits where a user spent at least 15 minutes and for each visit, also recorded its duration. 100,000 of the extracted location visits in downtown Houston were sampled to form the dataset used in our experiments, which contains three columns: latitude, longitude and visit duration. We let visit duration to be the measure attribute. 


\noindent\textbf{Query Distribution}. Our experiments consider query functions consisting of average, sum, standard deviation and median aggregation attributes together with two different predicate functions. First, similar to \cite{ma2019dbest}, our experiments show the performance on the predicate function defined by the WHERE clause in Sec.\ref{sec:def}. We consider up to 3 active attributes in the predicate function. To generate a query instance with $r$ active attributes, we first select, uniformly at random, $r$ activate attributes (from a total of $d$ possible attributes). Then, for the selected active attributes, we randomly generate a range. Unless otherwise stated, the range for each active attribute is uniformly distributed. This can be thought of as a more difficult scenario for {NeuroDB} as it requires approximating the query function equally well over all its domain, while also giving a relative advantage to other baselines, since they are unable to utilize the query distribution. Unless otherwise stated, for all dataset except Veraset, we report the results for one active attributes and use the average aggregation function. For Veraset, we report the results setting latitude and longitude as active attributes. Second, to show how NeuroDB can be applied to application specific RAQs, in Sec.\ref{exp:veraset}, we discuss answering the query of median visit duration given a \textit{general} rectangle on Veraset dataset. 


\noindent\textbf{Measurements}. In addition to query time and space used, we report the normalized absolute error for a query in the set of test queries, $T$, defined as $\frac{|f_D(q)-\hat{f}_D(q, \theta)|}{\frac{1}{|T|}\sum_{q\in T}|f_D(q)|}$. The error is normalized by average query result magnitude to allow for comparison over different data sizes when the results follow different scales. 

\begin{table}[t]
\hspace{-0.5cm}
\begin{minipage}{0.5\columnwidth}
    \centering
    \begin{tabular}{c|c|c}
        \textbf{Dataset} & \textbf{\# Points} &\textbf{Dim}. \\\hline
        \begin{tabular}{@{}c@{}} G5, G10\\ G20 \end{tabular}
          & 100,000 & \begin{tabular}{@{}c@{}} 5, 10\\ 20 \end{tabular} \\\hline
        PM2.5 \cite{liang2015assessing} & 41,757 & 4 \\\hline
        TPC1 \cite{nambiar2016the} & 2,653,123 & 13 \\\hline
        TPC10 \cite{nambiar2016the} & 26,532,166 & 13\\\hline
        Veraset (VS)  & 100,000 & 3
    \end{tabular}
    \caption{Datasets for range aggregate queries}
    \label{tab:dataset_size}
\end{minipage}
\hspace{0.1cm}
\begin{minipage}{0.5\columnwidth}
    \centering
    \begin{tabular}{c|c|c}
        \textbf{Dataset} & \textbf{\# Points}  & \textbf{Dim.} \\\hline
        \begin{tabular}{@{}c@{}}GloVe (GV)\\ \cite{pennington2014glove}  \end{tabular}  & 1,191,887 &
        \begin{tabular}{@{}c@{}}25, 50 \\ 100, 200  \end{tabular}
        \\\hline
        GIST \cite{gist} & 1,000,000 & 960 \\\hline
        KDD \cite{kdd} & 1,009,745 & 36 \\\hline
        IPUMS \cite{ipums} & 70,187  & 60
    \end{tabular}
    \caption{Datasets for distance to nearest neighbour query}
    \label{tab:nn_dataset_size}
\end{minipage}
\end{table}

\begin{figure*}[t]
    \begin{minipage}[t]{.49\textwidth}
        \centering
    	\includegraphics[width=\textwidth]{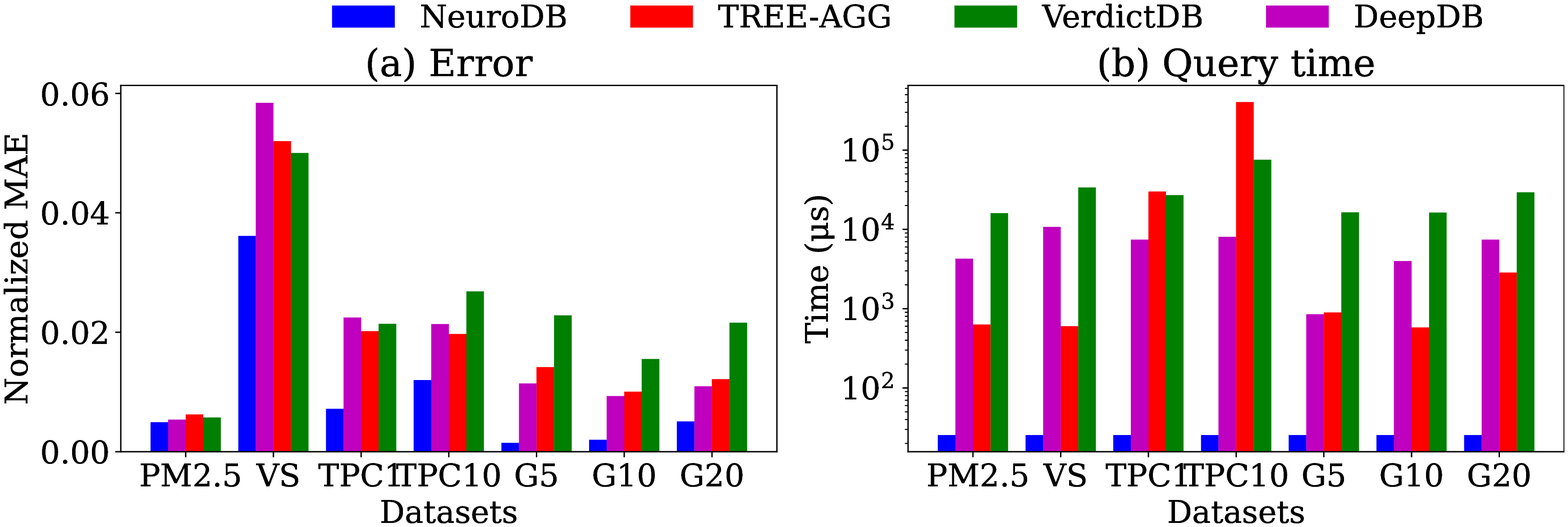}
    	    	\vspace{-0.65cm}
        \caption{RAQs on different datasets}
        \label{fig:exp:range_agg_data}
    \end{minipage}
    \hfill
    \begin{minipage}[t]{.49\textwidth}
        \centering
        \includegraphics[width=1\textwidth]{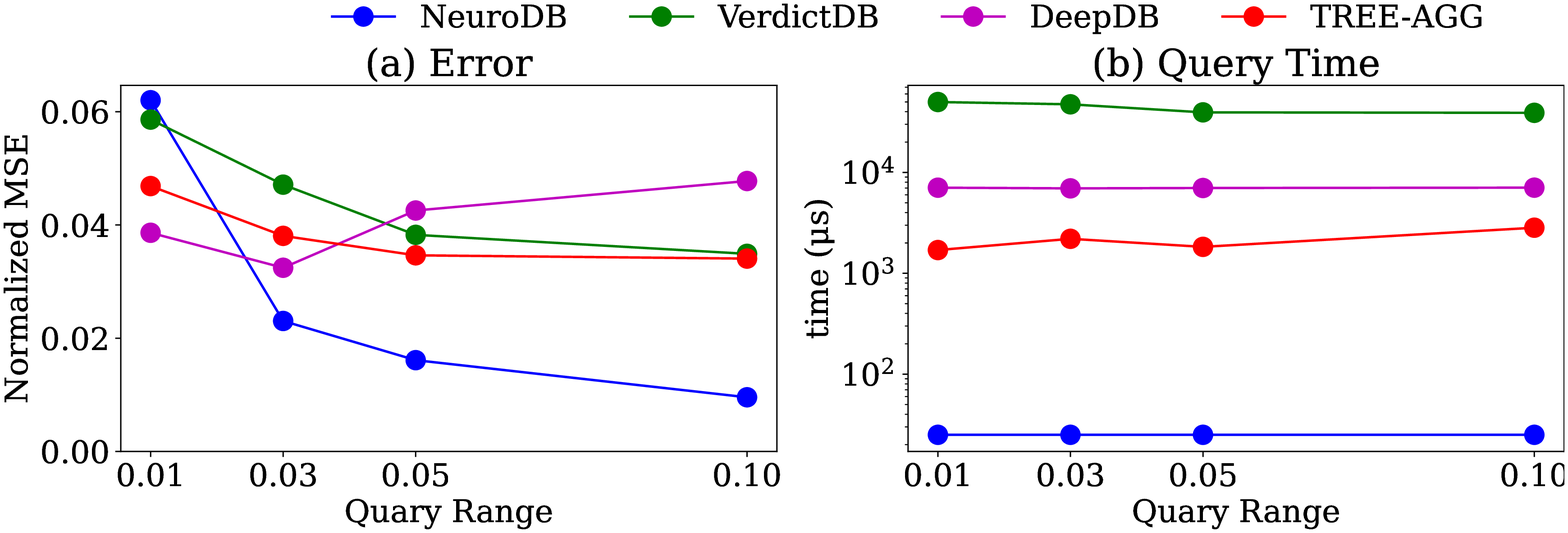}
        \vspace{-0.65cm}
        \caption{Impact of query range on RAQs}
        \label{fig:exp:range_agg_qrange}
    \end{minipage}
\end{figure*}
\begin{figure*}[t]
\vspace{-0.1cm}
    \begin{minipage}[t]{.49\textwidth}
        \centering
    	\includegraphics[width=1\textwidth]{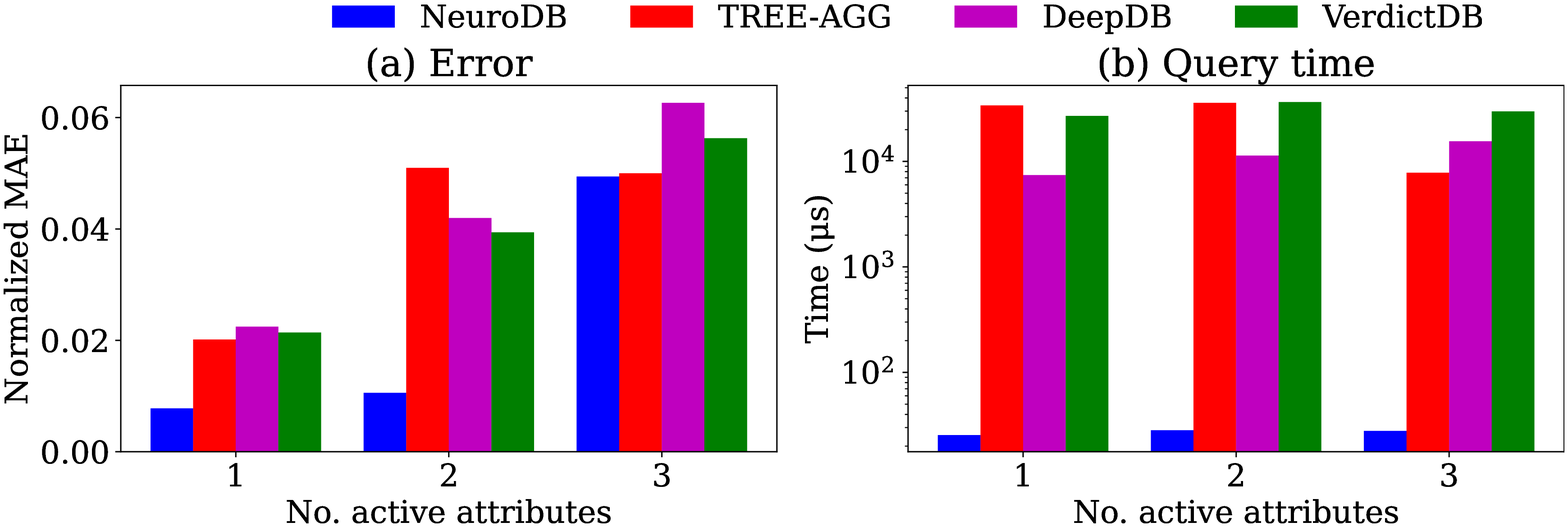}
    	\vspace{-0.65cm}
        \caption{Impact of number of active attributes on RAQ}
        \label{fig:exp:range_agg_dim}
    \end{minipage}  
    \hfill
    \begin{minipage}[t]{.49\textwidth}
        \centering
        \includegraphics[width=\textwidth]{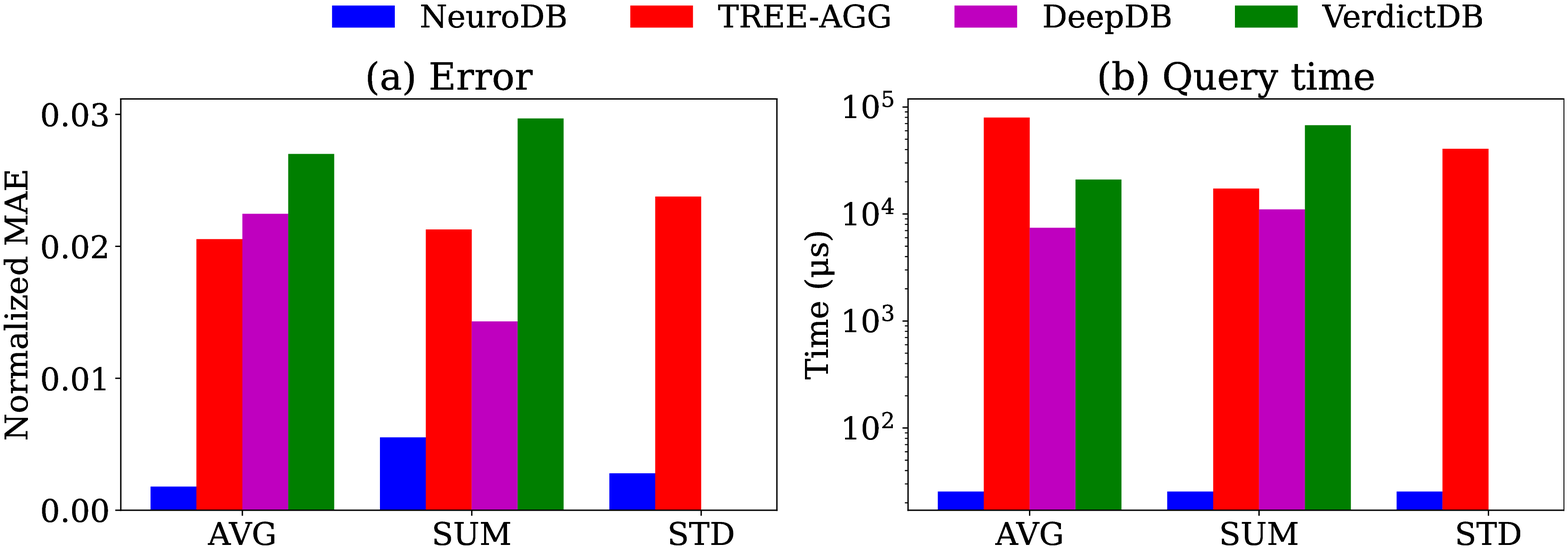}    	
        \vspace{-0.65cm}
        \caption{Impact of agg. function on RAQs}
        \label{fig:exp:range_agg_aggf}
    \end{minipage}
\end{figure*}

\noindent\textbf{Baseline Algorithms}. We considered DBEst \cite{ma2019dbest} and DeepDB \cite{hilprecht2019deepdb} as the state-of-the-art model-based approximate query processing engines. Both algorithms learn models of the data and answer specific queries based on the models of the data. We use the open-source implementation of DBEst available at \cite{dbest_imp} and DeepDB at \cite{deepdb_imp}. In a preliminary experiment, we compared the performance of DBEst and DeepDB on TPC1, for queries with one active attribute and observe that DeepDB performs much better than DBEst. Furthermore, DBEst's implementation does not support multiple active attributes. Thus, we chose DeepDB as our model-based baseline. We use the default parameter setting for DeepDB. Furthermore, we use VerdictDB \cite{park2018verdictdb} as our sampling-based baseline, using its publicly available implementation \cite{verdict_imp}. We also implemented a sampling-based baseline designed specifically for range aggregate queries, referred to as TREE-AGG. In a pre-processing step and for a parameter $k$, TREE-AGG samples $k$ data points from the database uniformly. Then, for performance enhancement and easy pruning, it builds an R-tree index on the samples, which is well-suited for range predicates. At query time, by using the R-tree, finding data points matching the query is done efficiently, and most of the query time is spent on iterating over the points matching the predicate to compute the aggregate attribute required. For both TREE-AGG and VerdictDB, we set the number of samples so that the error is similar to that of DeepDB, as we assume the-state-of-the-art algorithm, DeepDB, answers the queries with an acceptable error rate.


\subsubsection{Results Across Datasets} Fig.~\ref{fig:exp:range_agg_data}  (a) shows the error on different datasets, where NeuroDB provides a lower error rate than the baselines. Fig.~\ref{fig:exp:range_agg_data} (b) shows that NeuroDB achieves this while providing multiple orders of magnitude improvement in query time. NeuroDB has a relatively constant query time. This is because, across all datasets, NeuroDB's architecture only differs in its input dimensionality, which only impacts number of parameters in the first layer of the model and thus changes model size by very little. Due to our use of small neural networks, we observe that model inference time for NeuroDB is very small and in the order of few microseconds, while the modeling choices of DeepDB leads to query time often multiple orders of magnitude larger. Furthermore, the results on G5 to G20 show the impact of data dimensionality on the performance of the algorithms. We observe that, for NeuroDB, the error increases as dimensionality increases, which is because the model needs to learn more information. The similar impact can be seen for DeepDB, manifesting itself in increased query time. Furthermore, the R-tree index of TREE-AGG often allows it to perform better than the other baselines, especially for lower dimensional data.  

\subsubsection{Different Workloads on TPC1}
\hfill\\\noindent\textbf{Impact of Query Range}. We experimented with setting the query range to $x$ percent of the domain range, for $x\in\{1, 3, 5, 10\}$, presented in Fig.~\ref{fig:exp:range_agg_qrange}. We observe that error rate of NeuroDB increases for smaller query ranges since for smaller ranges NeuroDB needs to memorize where exactly each data point, rather then learning the overall distribution of data points which can be done for larger ranges. Nevertheless, NeuroDB provides better accuracy than the baselines for query ranges at least 3 percent, while performing queries orders of magnitude faster for all ranges. We note that if more accurate answers are needed for smaller ranges in an application, increasing the model size of NeuroDB can improve its accuracy at the expense of query time (as studied in Sec.~\ref{exp:veraset}).  

\noindent\textbf{Impact of no. of active attributes on Query Time}.  We vary the number of active attributes in the range predicate from one to three. The accuracy for all the algorithms drops when there are more active attributes, with NeuroDB outperforming the algorithms both in accuracy and query time. For NeuroDB, the drop in accuracy can be due to larger training size needed for higher dimensional data to achieve the same accuracy, while our training procedure is the same for all dimensions. 

\noindent\textbf{Impact of Aggregation Function}. Fig.~\ref{fig:exp:range_agg_aggf} shows how different aggregation functions impact performance of the algorithms. Overall, NeuroDB is able to outperform the algorithms for all aggregation functions. VerdictDB  and DeepDB implementation did not contain the implementation for STD and thus their results are not reported for that aggregation function. 

\begin{figure*}[t]
\vspace{-0.15cm}
    \begin{minipage}[t]{.49\textwidth}
        \centering
    	\includegraphics[width=1\textwidth]{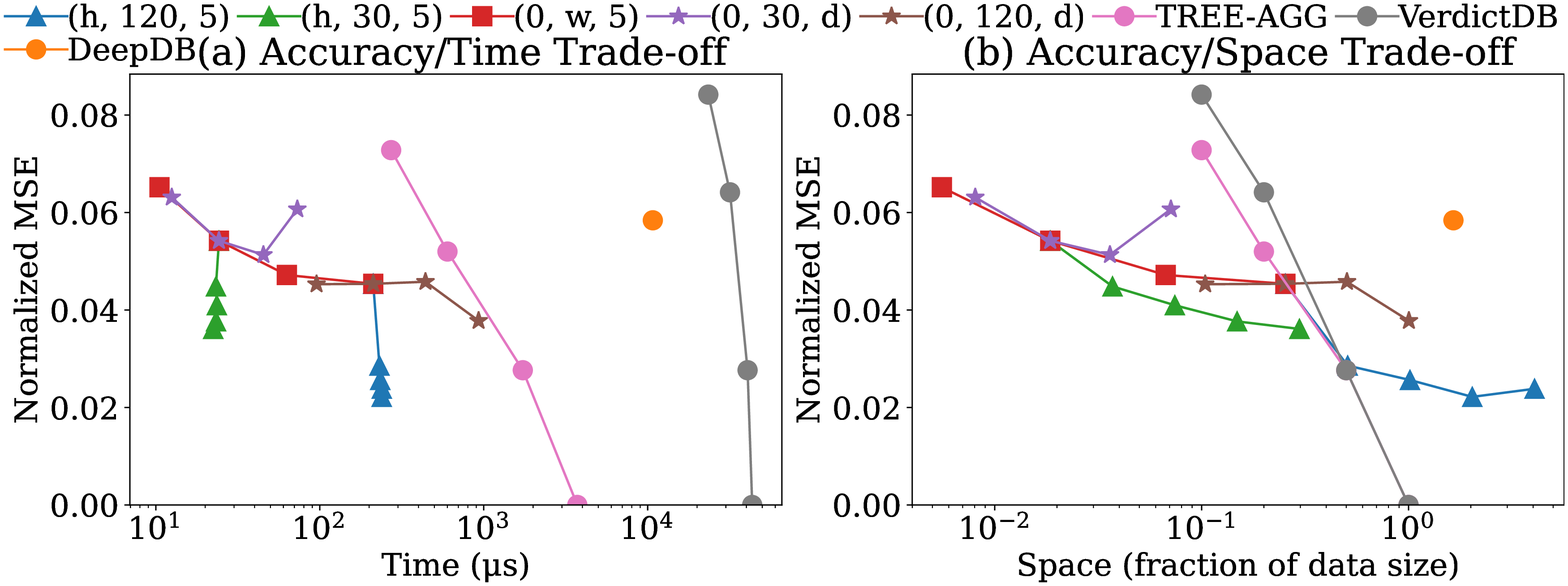}    	    	
    	\vspace{-1.2cm}
        \caption{Time/Space/Accuracy Trade-Off on Veraset}
        \label{fig:exp:RAQ_tradeoff}
    \end{minipage}
    \hfill
    \begin{minipage}[t]{.49\textwidth}
        \centering
    	\includegraphics[width=1\textwidth]{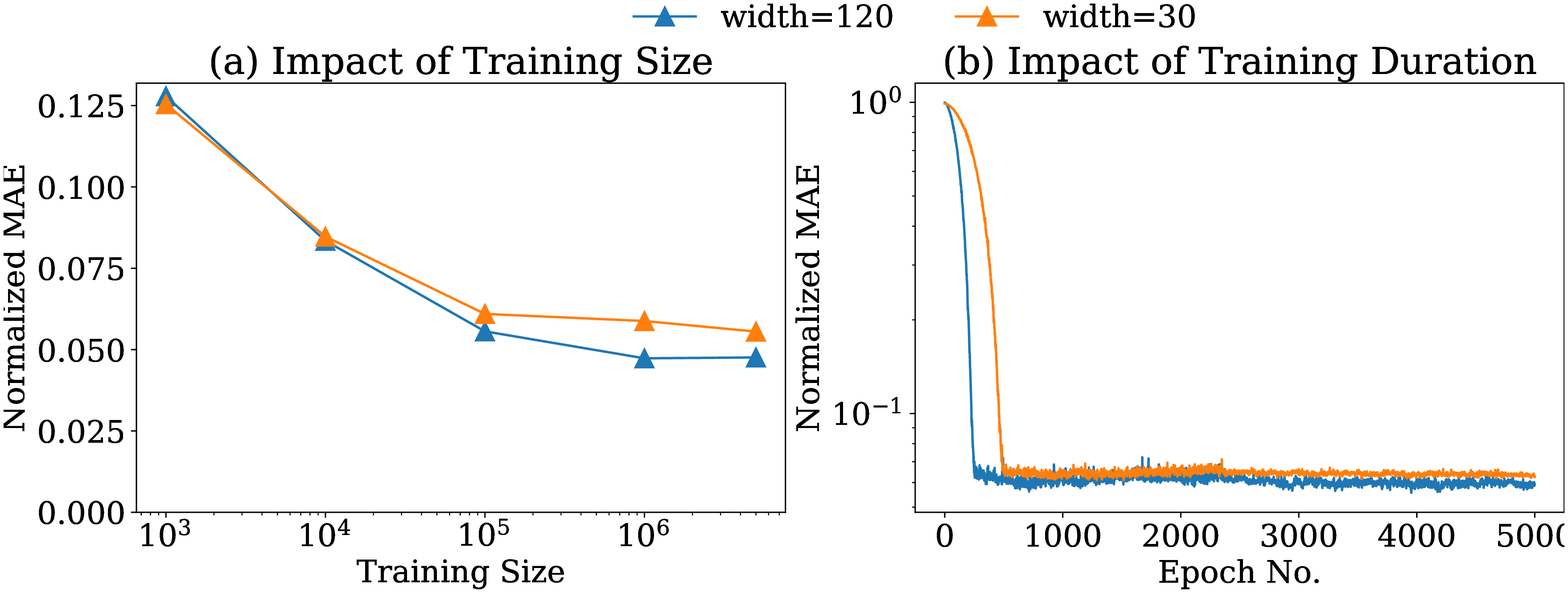}
    	\vspace{-1.2cm}
        \caption{Impact of training size and duration}
        \label{fig:exp:RAQ_training}
    \end{minipage}
\end{figure*}

\begin{figure}[t]
\vspace{-0.3cm}
    \centering
    \begin{minipage}{0.47\columnwidth}
    \includegraphics[width=\textwidth]{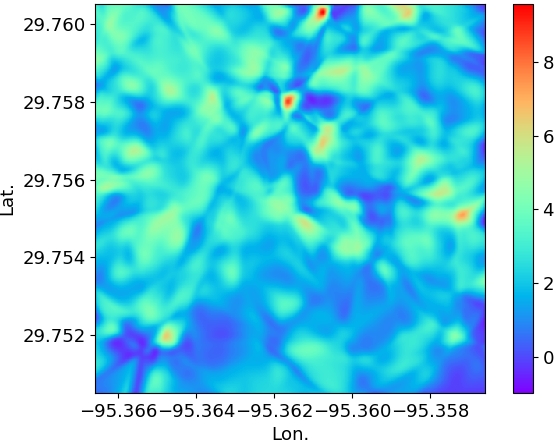}
    \end{minipage}
    \hfill
    \begin{minipage}{0.47\columnwidth}
    \includegraphics[width=\textwidth]{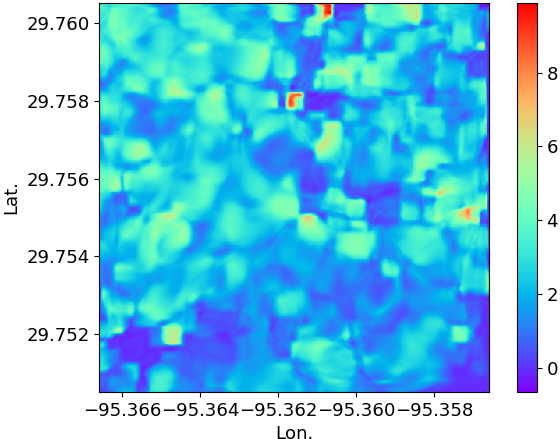}
    \end{minipage}
    \caption{Learned NeuroDB with depth 5 (left) and 10 (right) for the avg. visit duration example. Color shows the visit duration in hours.}
    \label{fig:learned_query_functions}
\end{figure}

\subsubsection{Results on Veraset}\label{exp:veraset}\hfill\\
\noindent\textbf{Time/Space/Accuracy Trade-Offs}. We study different time /space/accuracy trade-offs achievable by NeuroDB and other methods, shown in Fig.~\ref{fig:exp:RAQ_tradeoff}. For NeuroDB, we vary number of layers (referred to as depth of the neural network), $d$, number of units per layer (referred to as width of the neural network), $w$, and height of the kd-tree, $h$, to see their impact on its time/space/accuracy. Fig.~\ref{fig:exp:RAQ_tradeoff} shows several possible combinations of the hyperparameters. For each line in Fig.~\ref{fig:exp:RAQ_tradeoff}, NeuroDB is run with two of those hyperparameters kept the same and another one changing. Labels of the lines can be interpreted as follows. The line labels are of the form $(height, width, depth)$, where two of \textit{height, width} or \textit{depth} have numerical values and are the constant hyperparameters for that particular line. Furthermore, the value of one of \textit{height}, \textit{width} or \textit{depth} is $\{d, w, h\}$ and is the variable hyperparameter for the plotted line. For example, line labelled (h, 120, 5) means the experiments for the corresponding line are with a NeuroDB architecture with 120 number of units per layer, 5 layers and each point plotted corresponds to a different value for the kd-tree height, and label (0, 30, d) means the experiments are run with varying depth of the neural network, with kd-tree height 0 (i.e. only one partition) and the width of the neural network is 30. The hyperparameter values are as follows. For lines (h, 120, 5) and (h, 30, 50), kd-tree height is varied from 0 to 4, for the line labelled (0, w, 5) neural network width is in the set $\{15, 30, 60, 120\}$ and for lines labelled (0, 120, d) and (0, 30, d) neural network depth is in the set $\{2, 5, 10, 20\}$.

Finally, TREE-AGG and VerdictDB are plotted for sampling sizes of 100\%, 50\%, 20\% and 10\% of data size. For DeepDB only one result is reported since 
we observed in our experiments that changing the hyperparameters only impacts accuracy but not query time. Thus, we only report the result for the setting with the best accuracy.

Fig.~\ref{fig:exp:RAQ_tradeoff} (a) shows the trade-off between query time and accuracy. Overall, NeuroDB performs well when fast answers are required but some accuracy can be sacrificed, while if accuracy close to an exact answer is required, TREE-AGG can perform better. Furthermore, Fig.~\ref{fig:exp:RAQ_tradeoff} (b) shows the trade-off between space consumption and accuracy. Similar to time/accuracy trade-offs, we observe that when the error requirement is not too stringent, NeuroDB can answer queries by taking a very small fraction of data size. Finally, NeuroDB outperforms DeepDB in all the metrics. Finally, comparing TREE-AGG with VerdictDB shows that, on this particular dataset, the sampling strategy of VerdictDB does not improve upon uniform sampling of TREE-AGG while the R-tree index of TREE-AGG improves the query time by orders of magnitude of VerdictDB.

Moreover, Fig~\ref{fig:exp:RAQ_tradeoff} shows the interplay between different hyperparameters of NeuroDB. We can see that increasing depth and width of the neural networks improves the accuracy, but after a certain accuracy level the improvement plateaus and accuracy even worsens if depth of the neural network is increased but the width is too small (i.e., the purple line). Nevertheless, using our kd-tree partitioning method allows for further improving the time/accuracy trade-off as it improves the accuracy at almost no cost to query time. We also observe that kd-tree improves the space/accuracy trade-off, compared with increasing the width or depth of the neural networks. This shows that our paradigm of query specialization is in fact beneficial, as learning multiple specialized models each for a different part of the query space performs better than learning a single model for the entire space.

\noindent\textbf{Median visit duration query function}. Here, we consider the query of median visit duration given a \textit{general} rectangular range. Specifically, we consider the predicate function that takes as its input coordinates of two points $p_1$ and $p_2$, that represent the location of two non-adjacent vertices of the rectangle, and an angle, $\phi$, that defines the angle the rectangle makes with the x-axis. Then, given $q=(p_1, p_2, \phi)$, the query function is to return median of visit duration of records falling in the rectangle defined by $q$. This is a common query for real-world location data, and data aggregators such as SafeGraph \cite{safegraph} publish such information.

Neither DeepDB nor DBEst can answer this query, since the predicate function is not supported by those methods, and extending those methods to support them is not trivial. On the other hand, NeuroDB can readily be used to answer this query function. Although VerdictDB can be extended to support this query function, the current implementation does not support the aggregation function, so we do not report the results on VerdictDB. Table \ref{tab:median_visit_res} shows the results for this query function. Overall, the performance of the methods is similar to other results on Veraset dataset, reported in Fig.~\ref{fig:exp:RAQ_tradeoff}.

\begin{table}[t]
    \centering
    \begin{tabular}{c|c|c|c|c}
        Metric &  NeuroDB  & TREE-AGG & DeepDB & VerdictDB \\\hline
        Norm. MAE & 0.045 & 0.052 & N/A & N/A \\\hline
        Query time ($\mu s$) & 25 & 601 & N/A & N/A
    \end{tabular}
    \vspace{1pt}
    \caption{Median visit duration for general rectangles}
    \label{tab:median_visit_res}
\end{table}

\noindent\textbf{Impact of Training Size and Duration}.  Fig.~\ref{fig:exp:RAQ_training} shows the performance of NeuroDB as number of training samples and number of epochs to train changes. The results are for a NeuroDB with tree height 0 (i.e., no partitioning), neural network depth 5 and with neural network widths of 30 and 120. In Fig.~\ref{fig:exp:RAQ_training} (a), we observe that at training size of about 100,000 sampled query points, both architectures achieve close to their lowest error and further increasing number of samples only marginally improves the performance. Furthermore, when sample size is small, both architectures perform similar to each other and the neural network with larger width only starts outperforming the smaller neural network when enough samples are available. Finally, Fig.~\ref{fig:exp:RAQ_training} (b) shows that both models converge in less than one thousand epochs, for training size $5\times 10^6$ and where each epoch contains 50 batches. To reach 5000 epochs, the total training time for neural network with widths 120 and 30 was 39.7 min and 37.15 min respectively.  

\noindent\textbf{Visualizing NeuroDB}. Fig.~\ref{fig:learned_query_functions} shows the function NeuroDB has learned for our running example, for two neural networks with the same architecture, but with depths 5 and 10. Comparing Fig.~\ref{fig:learned_query_functions} with Fig.~\ref{fig:query_functions}, we observe that NeuroDB learns a function with similar patterns as the ground truth but the sharp drops in the output are smoothened out. We also observe that the learned function becomes more similar to the ground truth as we increase the number of parameters. Note that the neural networks are of size about 9\% and 3.8\% of the data size. 


\begin{figure*}[t]
        \hspace{-0.8cm}
    \begin{minipage}[t]{.6\textwidth}
        \centering
    	\includegraphics[width=1\textwidth]{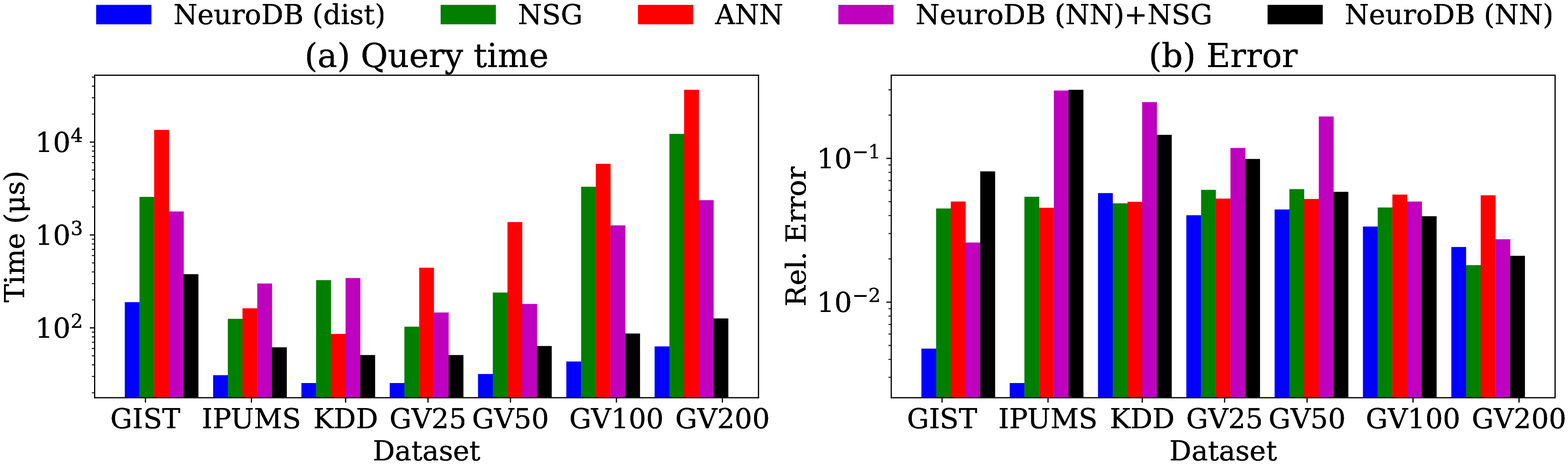}
    	   \vspace{-0.65cm}
    	\caption{Results across datasets for NN queries}
    	\label{fig:exp:dist_nn_datasets}
    \end{minipage}
    \begin{minipage}[t]{.45\textwidth}
        \centering
    	\includegraphics[width=1\textwidth]{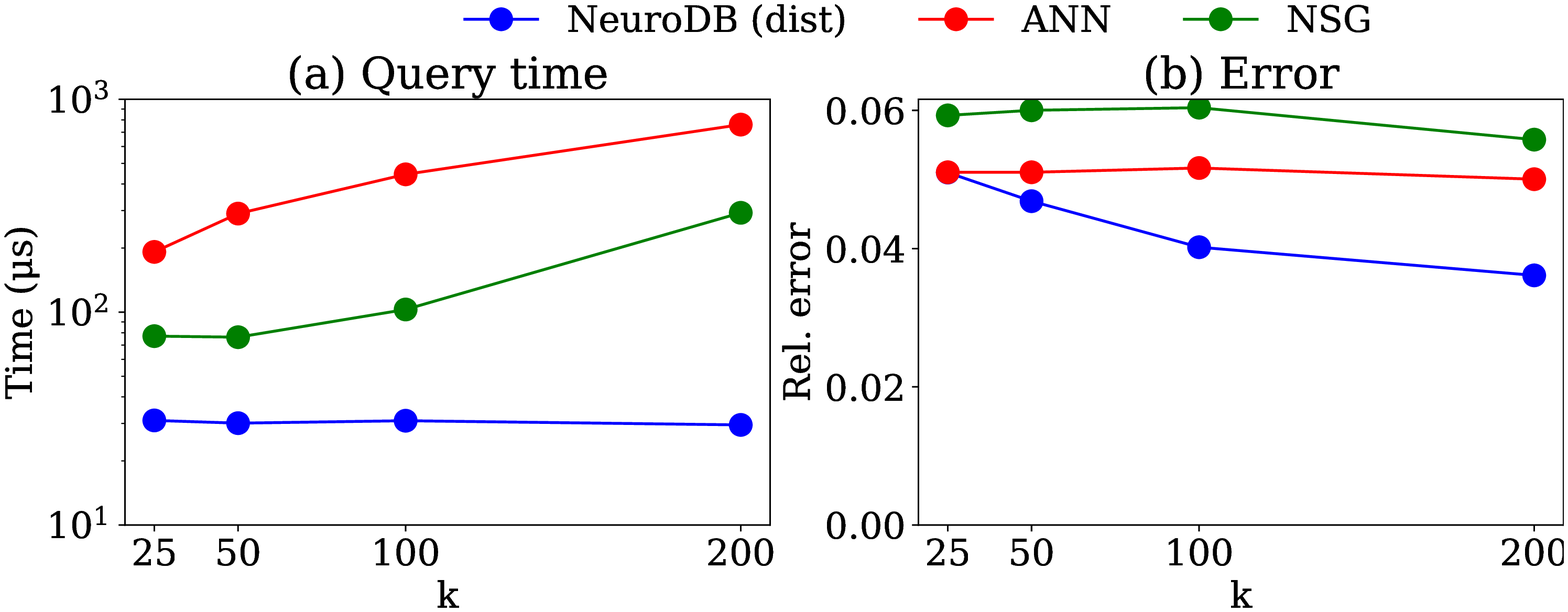}
    	\vspace{-0.65cm}
    \caption{Varying $k$ for dist. NN query}
    \label{fig:exp:dist_NN}
    \end{minipage}
\end{figure*}

\subsection{Distance to $k$-th Nearest Neighbour Query}\label{sec:exp:dist_NN}
\subsubsection{Setup}
In this section, we use the NeuroDB framework with the same architecture as for RAQs to answer distance to nearest neighbour queries as defined in Sec. \ref{sec:NN}. More results on hyper-parameter tuning for distance to nearest neighbour queries and a case study on nearest neighbour queries to provide more insight into what NeuroDB learns are provided in Secs.~\ref{appx:dist_exp} and \ref{appx:NN_exp}, respectively.  

\noindent\textbf{Datasets.} We use four common real datasets in our experiments whose size and dimensionality are reported in Table \ref{tab:nn_dataset_size}. For GloVe dataset, unless otherwise stated, we use the 25 dimensional version and refer by GV$<d>$ to the $d$ dimensional version of the dataset. 

\noindent\textbf{Query Distribution}. We assume the queries follow the data distribution, which can be true for the datasets considered. For instance, the GloVe dataset contains learned word representations. A distance to nearest neighbour query on this dataset can be interpreted as checking whether a set of words, $D$, contains a similar word to some query word $q$, or to check if a query $q$ would be an outlier in $D$ (e.g. to check whether a good representation for $q$ is learned or not). 

To generate the queries the points in the dataset are split into a training set of size $N_{train}$ and a testing set of size $N_{test}$ (points in the dataset are also training or testing queries). Unless otherwise stated, we set, $k=100$, $N_{test}=10,000$ and $N_{train}$ to the size of all the dataset except the test set. 

\noindent\textbf{Measurements}. We report time taken to answer a query, space used and average \textit{relative error}. Relative error, for the $k$-th nearest neighbour query $q$, when an algorithm returns the point $p_k$ while the correct answer is $p_k^*$ is defined as $\frac{|d(p_i, q)-d(p_i*, q)|}{d(p_i*, q)}$, where $d(x, y)$ is the Euclidean distance between $x$ and $y$. Furthermore, since the KDD dataset contained near duplicate records, the distance to the nearest neighbour could be very close to zero and relative error could be very large for small absolute error values. Thus, only for this dataset, we present the results for normalized mean absolute error (as defined in Sec.~\ref{sec:exp:rang_agg}) instead of relative error. 

\noindent\textbf{Baseline Algorithms}. We compare our algorithm with {NSG} \cite{fu2017fast} and {ANN} \cite{arya1998optimal}. {NSG} is a state-of-the-art graph-based algorithm, shown in \cite{fu2017fast} to outperform various existing methods. We use their publicly available implementation \cite{nsg_imp}. We also use {ANN} which is a wort-case optimal algorithm 
as implemented in ANN library \cite{ann_lib}. The algorithm, given a worst-case relative error parameter, $\epsilon$, returns $k$ nearest neighbours of $q$ such that the relative error for the $k$-th nearest neighbour in $D$ is at most $\epsilon$. For both methods to select algorithm parameters, we perform a grid search to find the parameters such that the query time is minimum while relative error is around 0.05.  


\subsubsection{Comparison Across Datasets}\label{sec:exp:dist_NN_res}
Fig.~\ref{fig:exp:dist_nn_datasets} shows the results on different datasets for both distance to $k$-NN and $k$-NN queries. Note that for the baseline algorithms, the error and query time for answering both $k$-NN and distance to $k$-NN is the same, since the algorithms, in both cases, return $k$ nearest neighbours as their output. In the figure, NeuroDB (dist) is NeuroDB trained to answer distance to $k$-NN queries and NeuroDB (NN) is trained to answer $k$-NN queries. Furthermore, we implemented NeuroDB (NN)+NSG, which is an algorithm that takes the output of NeuroDB (NN), and finds its first nearest neighbour in the database using NSG.  

\noindent\textbf{Distance to $k$-NN}. First, observe that NeuroDB is able to answer distance to $k$-NN queries orders of magnitude faster than the state-of-the-art while achieving similar accuracy. Furthermore, its query time is only marginally affected by data dimensionality (e.g., see GV25 to GV200 results) and changes very little for datasets with different sizes. We emphasize that these results are obtained without any parameter tuning for NeuroDB, and NeuroDB's architecture is exactly the same as when performing RAQs. This shows that NeuroDB can be applied to two different query types with minimal effort, and thus, can save time when designing a system.   

\noindent\textbf{$k$-NN}. Fig.~\ref{fig:exp:dist_nn_datasets} also shows the results for $k$-NN queries. Note that the output of NeuroDB (NN) is a $d$-dimensional point, but the point is not necessarily in the dataset. Thus, even though Fig.~\ref{fig:exp:dist_nn_datasets} shows that NeuroDB (NN) can answer $k$-NN queries with accuracy similar to NSG but much faster for high dimensional datasets, such an answer may not be useful for real-world applications. Thus, we investigated using NSG together with NeuroDB (NN). Such an algorithm first runs NeuroDB (NN), and then runs NSG to find the nearest neighbour of the output of NeuroDB (NN). Intuitively, this can improve query time because NeuroDB (NN) is very fast, and NSG only needs to find the first nearest neighbour instead of $k$-th nearest neighbour. We observe that, for high dimensional datasets, such an algorithm can improve the query time over standalone NSG while obtaining similar accuracy. Thus, such a hybrid approach can be useful, where NeuroDB is used as a heuristic to search the space, but we leave a full investigation of such hybrid approaches to the future work.

\subsubsection{Impact of $k$ on GloVe dataset}\label{sec:exp:dist_NN_glove}
Fig.~\ref{fig:exp:dist_NN} shows how changing $k$ impacts the performance. Overall $k$ has no impact on query time of {NeuroDB}, because for any value of $k$ the cost of using {NeuroDB} is just a forward pass of the neural network. However, {NSG} and {ANN} need to find all the $k$ nearest neighbours, and thus their performance deteriorates with $k$. Better support for larger values of $k$ is one of our motivations for using {NeuroDB}, as it can find the distance to $k$-th nearest neighbour without unnecessarily finding all $k$ nearest neighbours, which other methods do.

\section{Related Work}\label{sec:related_work}
Related works can be classified into 3 groups: algorithms for RAQs and nearest neighbour queries and machine learning methods for database queries.

\noindent\textbf{Algorithms for Approximate Query Processing}. Approximate query processing (AQP) has many applications in data analytics, with queries that contain an aggregation function and a selection predicate used to report statistics from the data. Broadly speaking, the methods can be divided into sampling-based methods \cite{hellerstein1997online, agarwal2013blinkdb, chaudhuri2007optimized, park2018verdictdb} and model-based methods \cite{graham2012synopses, schmidt2002propolyne, ma2019dbest, thirumuruganathan2019approximate, hilprecht2019deepdb}. Sampling-based methods use different sampling strategies  (e.g., uniform sampling, \cite{hellerstein1997online}, stratified sampling \cite{chaudhuri2007optimized, park2018verdictdb}) and answer the queries based on the samples.  Model-based methods develop a model of the data that is used to answer queries. The models can be of the form of histograms, wavelets, data sketches (see \cite{graham2012synopses} for a survey) or regression and density based models \cite{ma2019dbest, thirumuruganathan2019approximate, hilprecht2019deepdb}. Generally, these works follow two steps. First, a model of the data is created. Then, a method is proposed to use these data models to answer the queries. The important difference between these works and ours is that the models are created \textit{based on the database} to answer \textit{specific} queries. That is, a model is created that explains the data, rather than a model that predicts the query answer. For instance, regression and density based models of \cite{ma2019dbest} or the sum-product network of \cite{hilprecht2019deepdb} are models of the data that are created independent of potential queries. We experimentally showed that our modeling choice allows for orders of magnitude performance improvement. Secondly, specific models can answer specific queries, (e.g. \cite{ma2019dbest} answers only COUNT, SUM, AVG, VARIANCE,
STDDEV and PERCENTILE aggregations). However, our framework is query-type agnostic and can be applied to any aggregation function. In this respect, our approach is similar to sampling-based methods that can be applied to any aggregation function and selection predicate. However, sampling-based methods fail to capitalize on patterns available in either data points or query distribution, which results in worse performance. 

\noindent\textbf{Algorithms for Nearest neighbour query}. Nearest neighbour query has been studied for decades in the computer science literature \cite{friedman1975algorithm, guan1992equal, bei1985improvement}, and is a key building block for various applications in machine learning and data analysis \cite{shalev2014understanding, chandola2009anomaly, alfarrarjeh2020class, chen2005robust, yu2020combo}. For various applications, such as similarity search on images, it is important to be able to perform the query fast, a problem that becomes hard to address in high-dimensional spaces \cite{weber1998quantitative, gionis1999similarity, arya1998optimal}. As a result, more recent research has focused on approximate nearest neighbour query \cite{zheng2016lazylsh, huang2015query, wang2018randomized, fu2017fast, zhao2020song}. Generally speaking, all the methods iterate through a set of candidate approximate nearest neighbours and prune the candidate set to find the final nearest neighbours. The algorithms can be categorized into locality-sensitive hashing (LSH) \cite{gionis1999similarity, indyk1998approximate, zheng2016lazylsh, wang2018randomized, huang2015query}, product quantization \cite{jegou2010product, johnson2019billion}, tree-based methods  \cite{arya1998optimal, roussopoulos1995nearest},
and graph-based searching \cite{fu2017fast, hajebi2011fast, zhao2020song}. LSH-based and quantization-based methods map the query point to multiple buckets which are expected to contain similar points. 

Finding a small candidate set is difficult, and as dimensionality increases more candidate points need to be checked. However, our NeuroDB framework avoids accessing points altogether, and learns a function based on the data that can answer the queries accurately. Moreover, the size of the candidate set increases with $k$, for all the algorithms. That is, more points need to be searched when $k$ increases. However, in an application which only requires distance to the $k$-th nearest neighbour, NeuroDB can directly output the answer, without the value of $k$ affecting query time. We are unaware of any other work that studies distance to $k$-th nearest neighbour without finding any of the nearest neighbours.

\noindent\textbf{Machine Learning for Databases}. There has been a recent trend to replace different database components with learned models \cite{ortiz2018learning, hashemi2018learning, hilprecht2019deepdb, macke2018lifting, nathan2020learning, idreos2019learningb, idreos2019Learninga, kraska2019sagedb, mitzenmacher2018model, ma2019dbest, thirumuruganathan2019approximate, galakatos2018tree, galakatos2018tree}. Most of the effort has been in either indexing (on one dimensional indexes are studied in \cite{kraska2018case, galakatos2018tree, ding2019alex}, with extensions to multiple dimensions studied in \cite{nathan2020learning}, using Bloom filters in \cite{mitzenmacher2018model, macke2018lifting} and key-value stores in \cite{idreos2019learningb, idreos2019Learninga}) or approximate query processing (learning data distribution with a model \cite{thirumuruganathan2019approximate}, using reinforcement learning for query processing \cite{ortiz2018learning}, learning models based on the data to answer queries\cite{hilprecht2019deepdb, ma2019dbest} and cardinality estimation \cite{kipf2018learned, wu2021unified}) with \cite{kraska2019sagedb} discussing how they can be put together to design a database system. The main observation in these bodies of work is that a certain database operation (e.g. retrieving the location of a record for the case of indexing in \cite{kraska2018case}) can be replaced by a learned model. In our paper, we observe the overarching idea that answering \textit{any} query can be performed by a model, since any query is a function that can be approximated. In this respect, our work can be seen as a generalization of the recent work in this area. Solving this more general problem requires a learning method with strong representation power, which motivates our use of neural networks. This is in contrast with simpler models used in \cite{kraska2018case, kraska2019sagedb, ma2019dbest}. 


\vspace{-0.3cm}
\section{Conclusion}\label{sec:conclusion}
We introduced NeuroDB, a neural network framework for efficiently answering RAQs and distance to nearest neighbour queries, with orders of magnitude improvement in query time over the state-of-the-art algorithms.
We further showed that the same framework and neural network architecture can be used to answer both query types, showing the potential for utilizing a single framework to answer different query types and minimizing human time spent on designing algorithms for different query types. 
To improve NeuroDB, future work can focus on parallelism, better partitioning methods and understanding theoretical guarantees of the accuracy of the method. Model pruning methods \cite{blalock2020state} that remove some of the \textit{unimportant} model weights can also be considered to reduce model size and evaluation time for faster performance, at the cost of accuracy. Additionally, studying NeuroDB for dynamic data and changing query distribution can be an interesting future direction. A straightforward approach can be to frequently test NeuroDB on a (potentially changing) test set, and re-train the neural networks whose accuracy fall below a certain threshold, but it may be possible to update the neural networks more cleverly. The general problem of updating a neural network is studied in the machine learning literature and is an active area of research \cite{quionero2009dataset, long2017deep, wang2019transferable, liu2019transferable}. However, interesting problems arise in the case of NeuroDB, since insertion of a new data point changes the query function in a specific and query dependent way. 

\end{sloppy}
\bibliographystyle{IEEEtran}
\bibliography{references}

\clearpage
\appendix
\subsection{Parallelism}\label{appx:parallel}
An important aspect of using neural networks is the ability to perform matrix multiplications in parallel. In this paper, to allow for fair comparison with existing non-parallelizable methods, we discuss our results in a single threaded model and perform our experiments on CPUs. Meanwhile, parallelizing NeuroDB has its own challenges. The forwarded passes on our networks consists of only one query, thus, no batching is performed at query time. Furthermore, each neural network used is relatively small. As a result, a lot of synchronization needs to be done among different threads during the forward pass (we note that performing nearest neighbour query in the batch setting is also studied in the literature \cite{johnson2019billion}, in which parallelization becomes easier for our approach, with less synchronization needed). For instance, assigning a thread to each perceptron means output of each core needs to be sent to all other cores at every layer, which slows down the forward pass. This is despite the fact this parallelization can theoretically reduce the time complexity by the number of cores. We believe more studies need to be conducted to reap the benefits of parallelization.



\begin{figure*}[t]
\centering
    \begin{minipage}[t]{.74\textwidth}
        \centering
    	\includegraphics[width=0.32\textwidth]{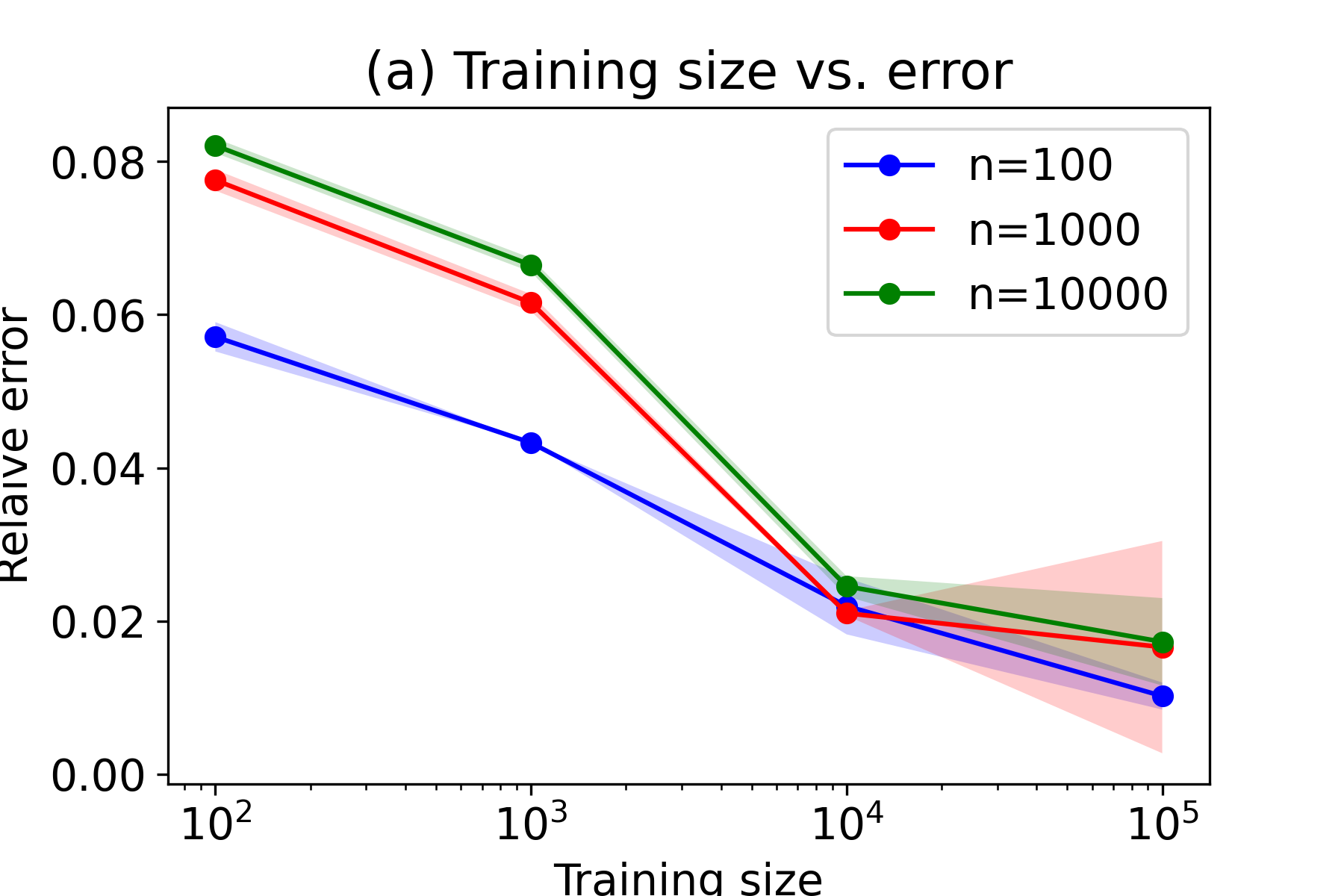}
    	\includegraphics[width=0.32\textwidth]{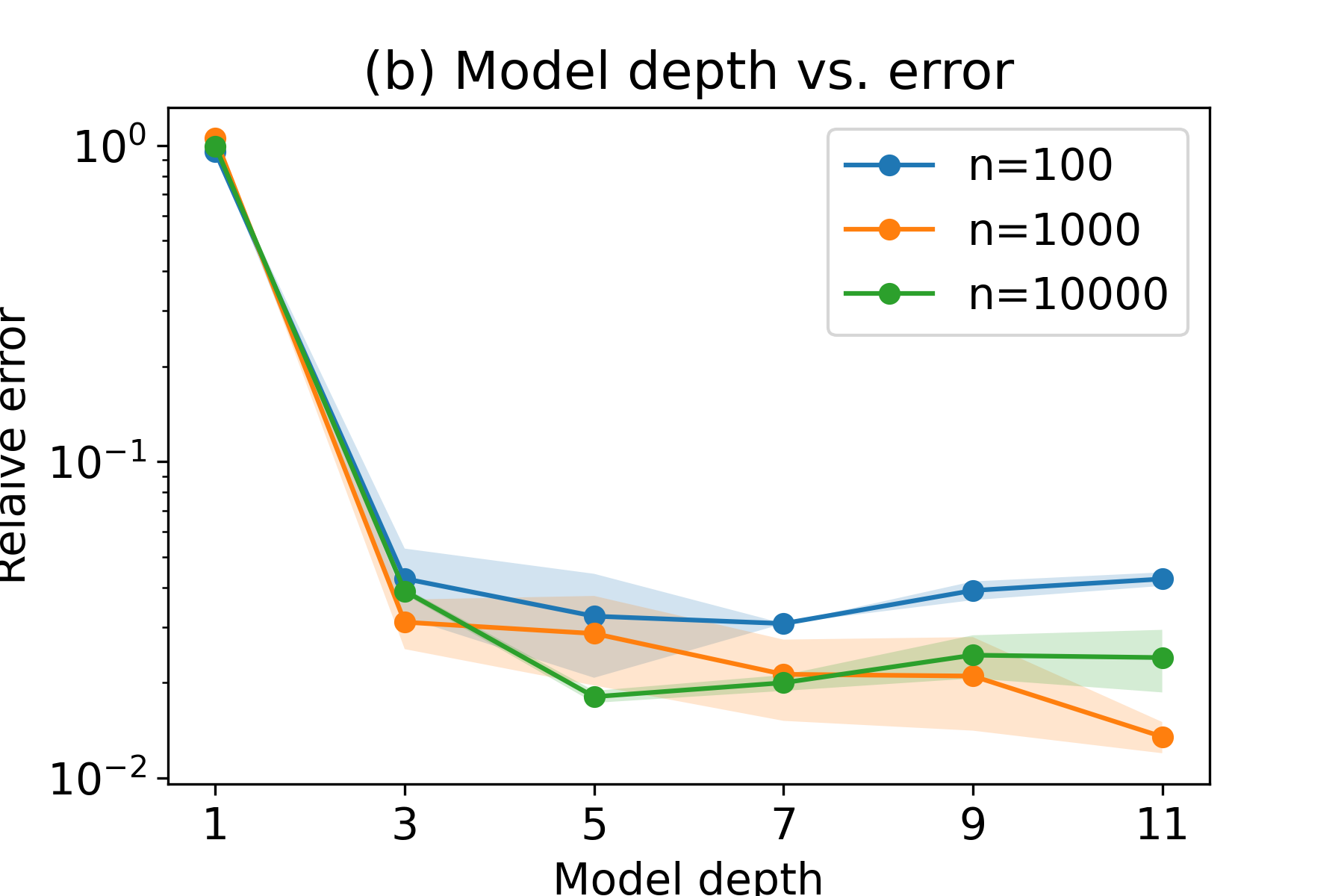}
    	\includegraphics[width=0.32\textwidth]{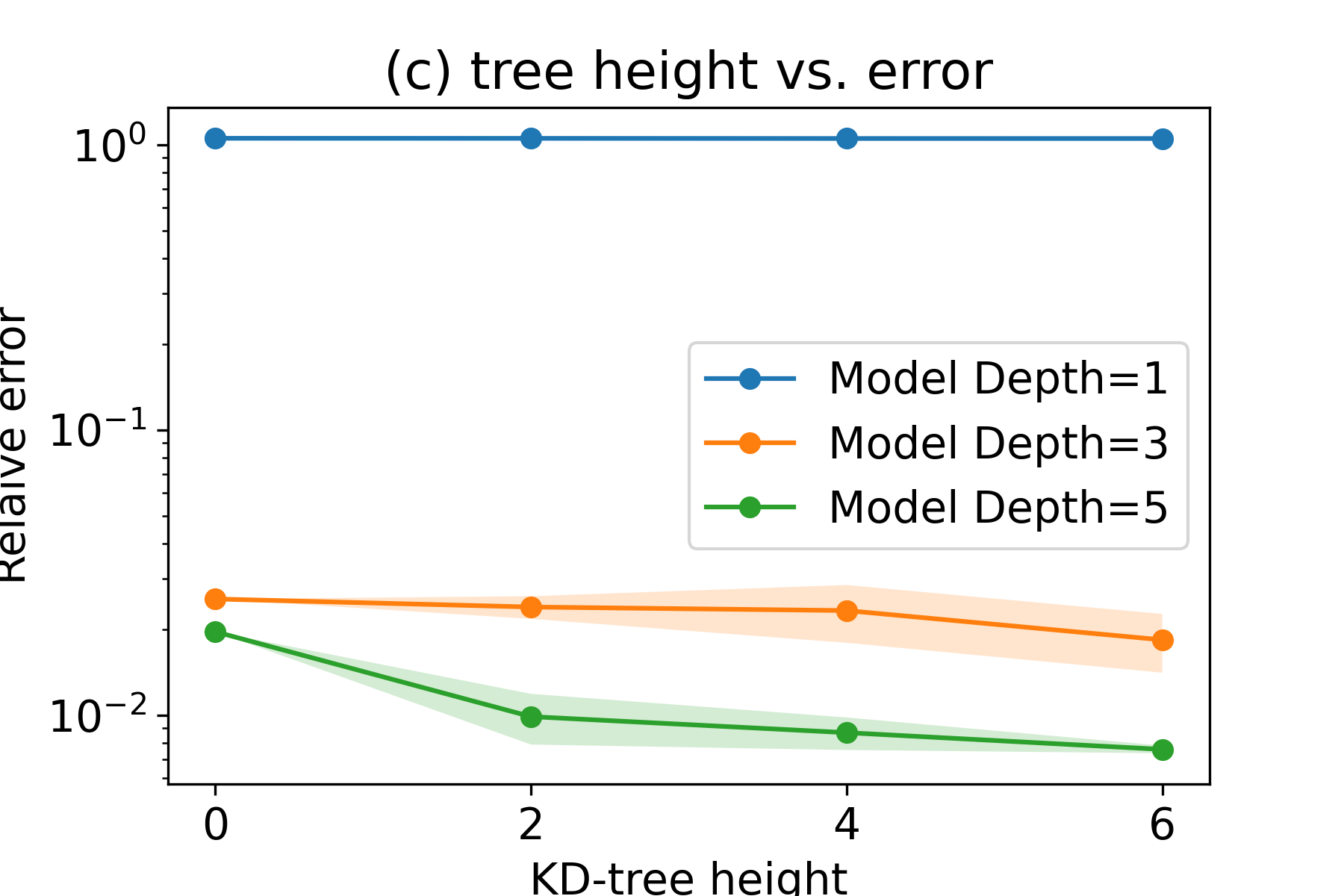}
    \caption{Impact of various hyperparameters on distance to nearest neighbour query}
    \label{fig:exp:dist_NN_params}
    \end{minipage}
        \begin{minipage}[t]{.23\textwidth}
           \centering
    	\includegraphics[width=\textwidth]{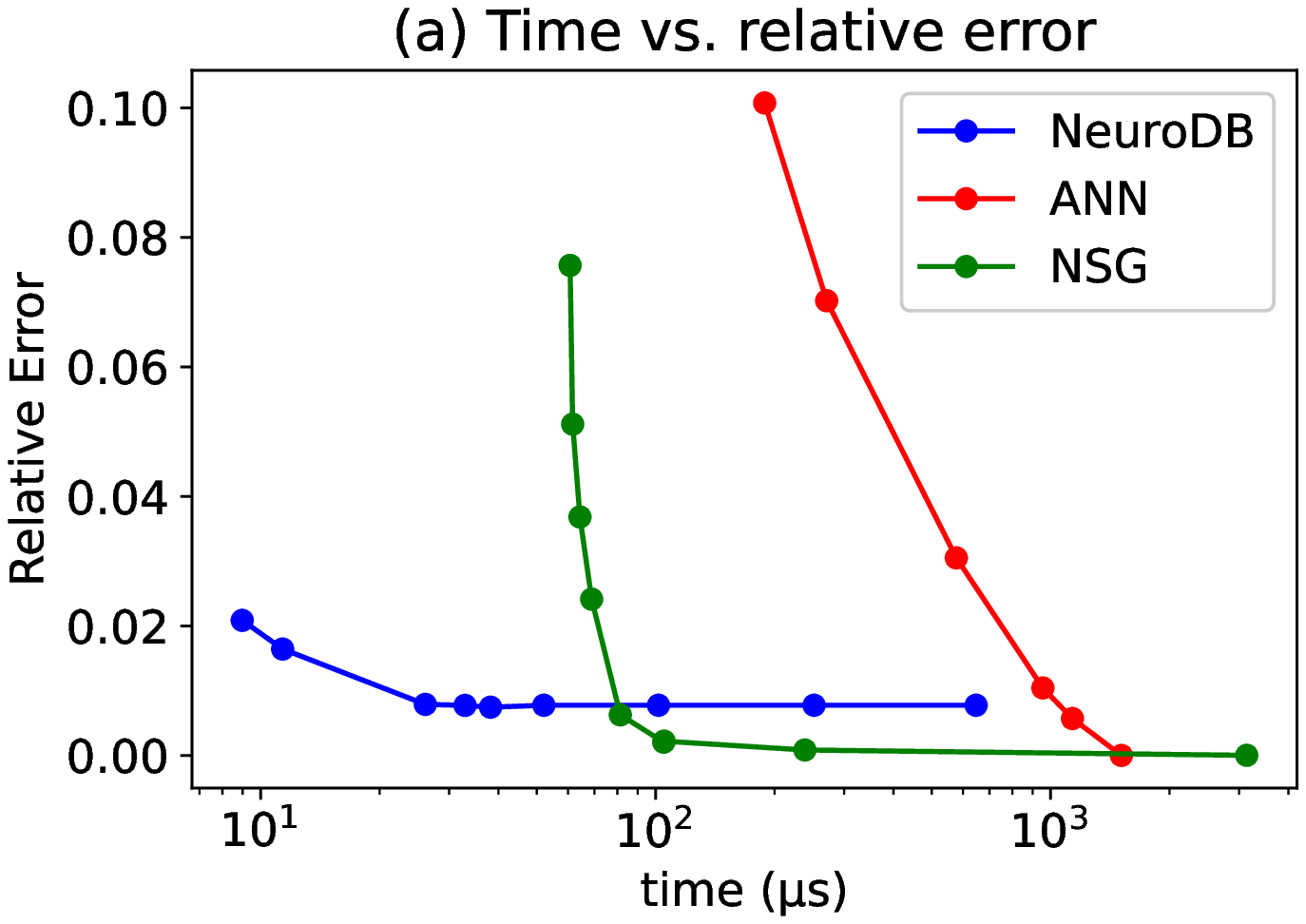}
    	\vspace{-0.5cm}
    	\caption{Accuracy/time trade-off}
    	\label{fig:exp:dist_nn_error_time}
    \end{minipage}
\end{figure*}
\subsection{Distance to $k$-NN Complementary Experimental Results}\label{appx:dist_exp}

\subsubsection{Impact of hyperparameters on accuracy}\label{sec:exp:hyperparam}
Experiments in this subsection are performed on uniform data and query distribution, with $n=10,000$, $d=20$ and $k=20$.

\noindent\textbf{Impact of Training Size.} Fig.~\ref{fig:exp:dist_NN_params} (a) shows the impact of training size on model accuracy. The results are average of three runs (three models are trained, where the randomness is due to the models being initialized randomly at the beginning of the training, as well as using SGD for training), and the shaded area shows the standard deviation. In this experiment, to keep the training time the same, the number of updates applied to the models, as well as the batch size, are kept the same across different training sizes (fewer epochs are run for larger training sizes, because each epoch contains more updates). We observe that as training size increases model accuracy improves. Furthermore, larger training size is more important for larger values of $n$. The increase in standard deviation for larger training sizes is due to the algorithm running for fewer epochs on larger training sizes.

\noindent\textbf{Impact of Model Depth}.  Fig.~\ref{fig:exp:dist_NN_params} (b) shows the impact of model depth on accuracy. First, we observe that a linear model (e.g. a neural network with only one layer) provides very poor accuracy, which justifies our use of deeper neural networks. Second, increasing model depth beyond a certain point does not necessarily improve accuracy. Increasing model depth can cause over-fitting, which explains the worsening of performance observed in the figure for larger model sizes. 

\noindent\textbf{Impact of Tree Height}.  Fig.~\ref{fig:exp:dist_NN_params} (c) shows the impact of the height of the kd-tree, which determines number of partitions used. We observe that increasing the height generally increases accuracy, but with larger models benefiting more. We note that in this experiment, for each partition, we keep the number of training samples used fixed (i.e., there are more training samples as more partitions are created). Overall, we observe that larger depth and more training size improve model accuracy. However, if the training size is fixed (e.g., if we don't have access to the query distribution), there is a limit to improvements obtained by increasing tree height, as  number of training samples per model will be reduced.

We also note that the standard deviation shown in Fig.~\ref{fig:exp:dist_NN_params} (c) is over different models in the NeuroDB and not multiple runs. The low standard deviation shows that all models responsible for different partitions obtain similar accuracy. 

\subsubsection{Accuracy/Time Trade-Off} 
Fig.~\ref{fig:exp:dist_NN} shows the accuracy/time trade-off of the algorithms on a subset of GV25 dataset. Specifically, the experiments here were run on 10,000 data points sampled from GV25. {NSG} and {ANN} are plotted at different error levels, and as can be observed query time increases as lower error is required. For {NeuroDB}, each point in the figure corresponds to a different neural network architecture. From left to right, query time of {NeuroDB} increases because a larger network architecture is used (we used a combination of increasing depth of the network as well as its width). We observe that, initially, as larger architectures are used, the ability of the model to learn increases and the accuracy improves. However, after a certain point, the model accuracy stops improving and even deteriorates. This can be attributed to two facts. First, as model size increases, it becomes more difficult to train the model (i.e., more training samples and more training iterations will be needed). Second, the model becomes more prone to over-fitting and may not perform well at evaluation time.

An interesting observation is that {NeuroDB} outperforms {NSG} and {ANN} in the low accuracy regime by an order of magnitude. However, after a certain accuracy level it becomes difficult to learn a {NeuroDB} that learns the query with that accuracy. Thus, the benefit of {NeuroDB} can be seen when fast answers are required, but some accuracy can be sacrificed. 

\begin{figure*}
    \centering
\begin{minipage}{0.3\textwidth}
    \centering
    \begin{tabular}{c c}
    \includegraphics[width=0.6\textwidth]{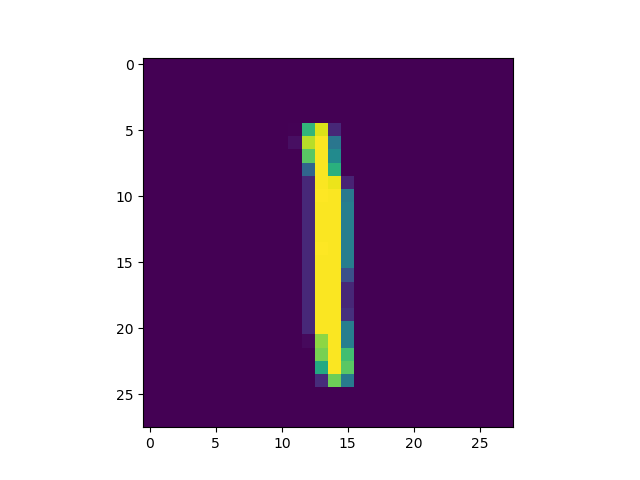} &
    \hspace{-1cm}\includegraphics[width=0.6\textwidth]{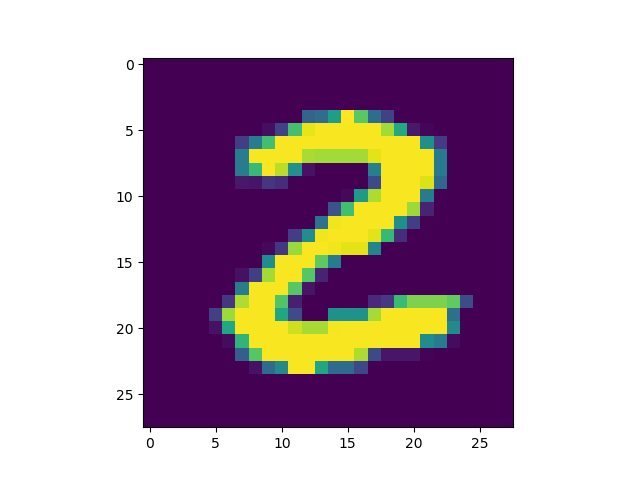} \\ \includegraphics[width=0.6\textwidth]{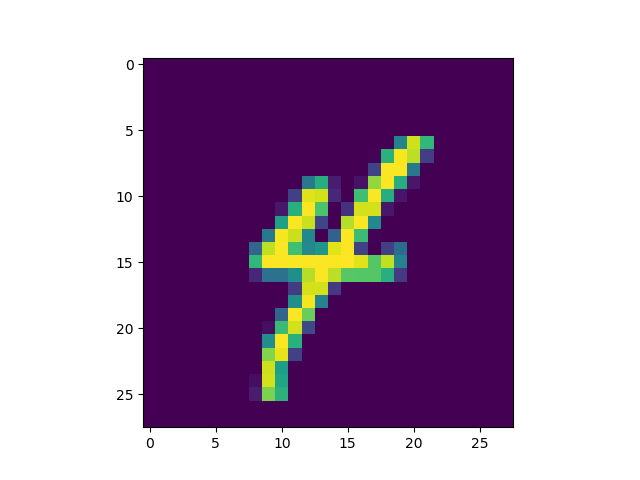} & \hspace{-1cm}\includegraphics[width=0.6\textwidth]{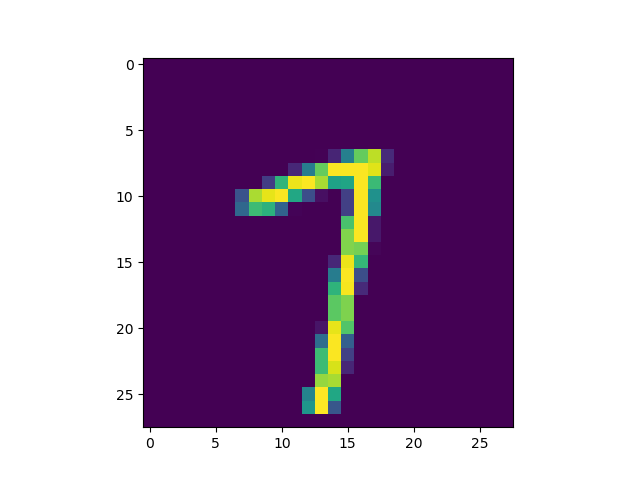} \\ \includegraphics[width=0.6\textwidth]{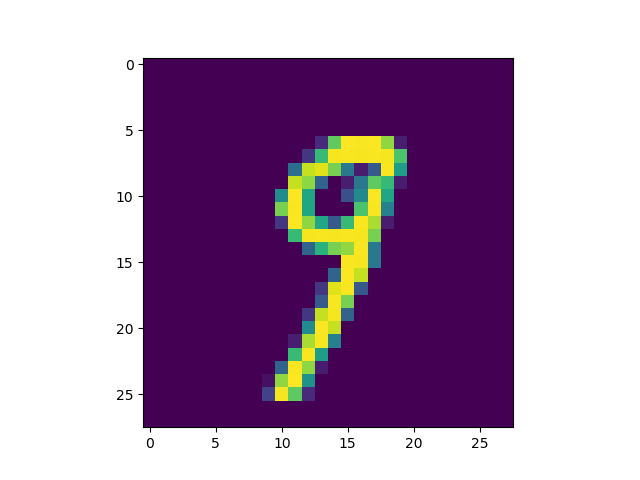} & 
    \end{tabular}
    \caption{Some of the images in the database}
    \label{fig:mnist_db}
\end{minipage}
\hfill
\begin{minipage}{0.65\textwidth}
    \centering
    \begin{tabular}{c c c c }
    \includegraphics[width=0.28\textwidth]{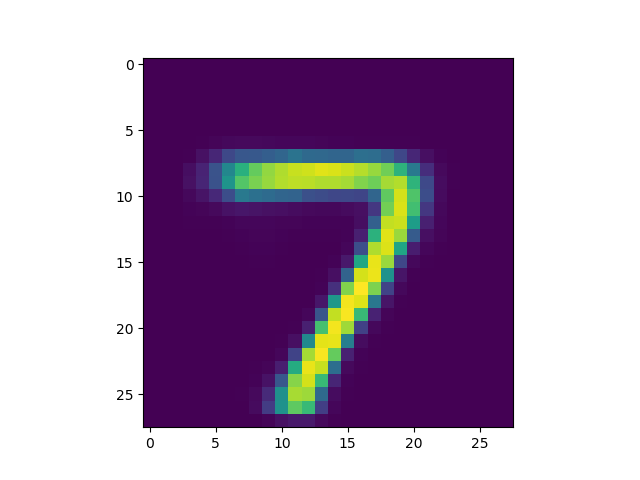} &
    \hspace{-1cm}\includegraphics[width=0.28\textwidth]{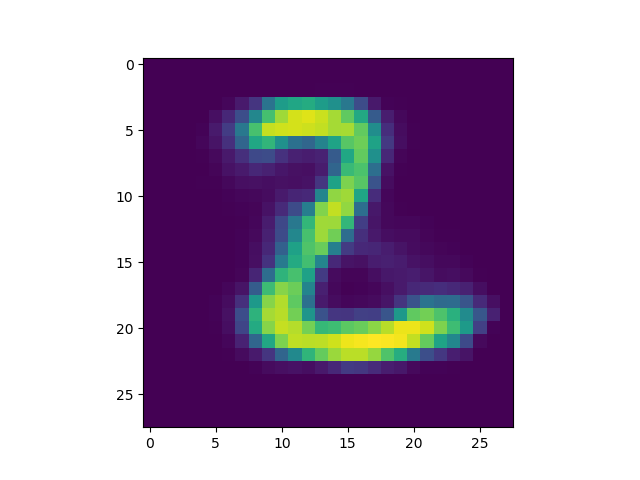} & \hspace{-1cm}\includegraphics[width=0.28\textwidth]{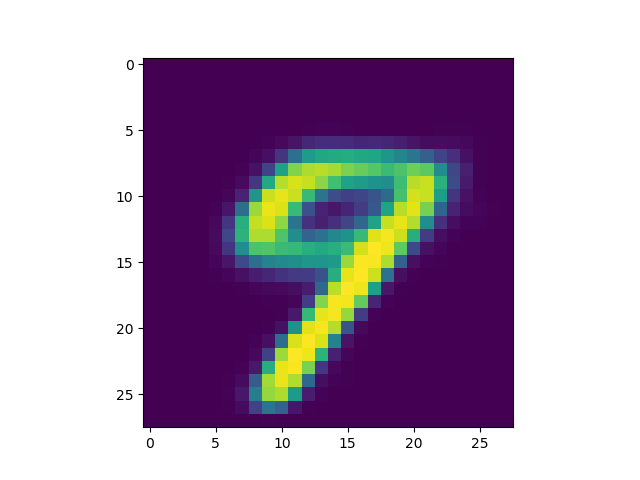} & \hspace{-1cm}\includegraphics[width=0.28\textwidth]{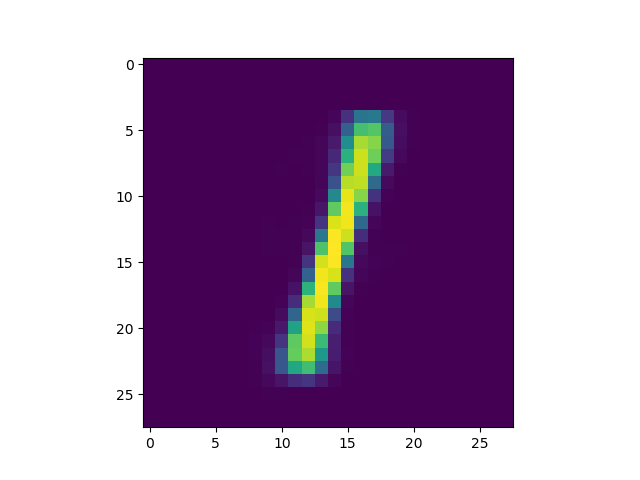} \\
    \includegraphics[width=0.28\textwidth]{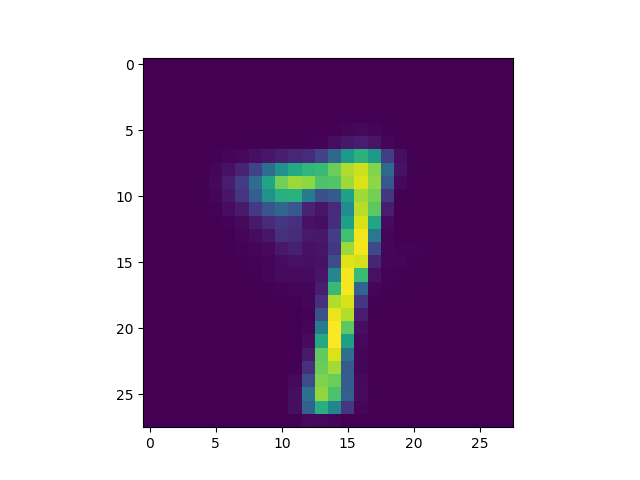} &
    \hspace{-1cm}\includegraphics[width=0.28\textwidth]{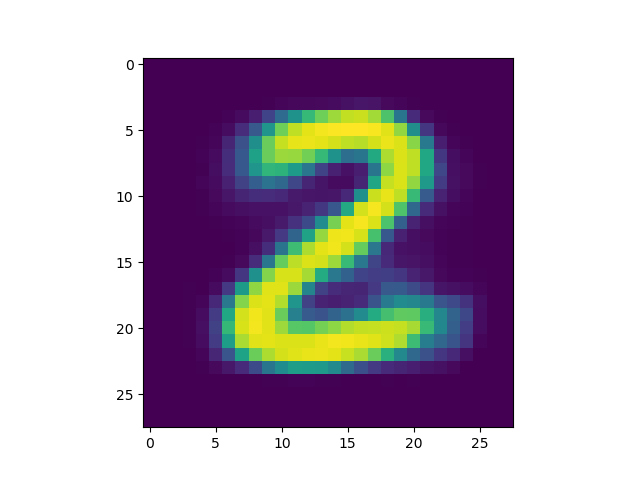} & \hspace{-1cm}\includegraphics[width=0.28\textwidth]{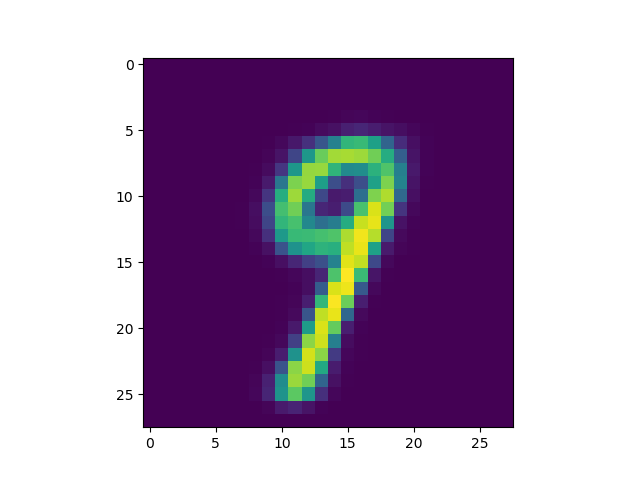} & \hspace{-1cm}\includegraphics[width=0.28\textwidth]{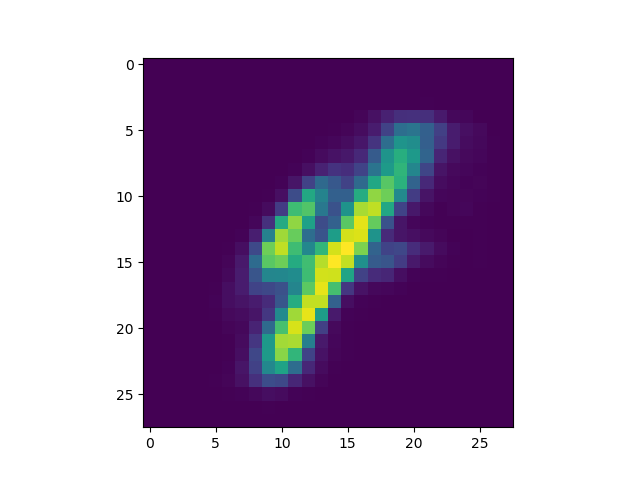} \\
    \includegraphics[width=0.28\textwidth]{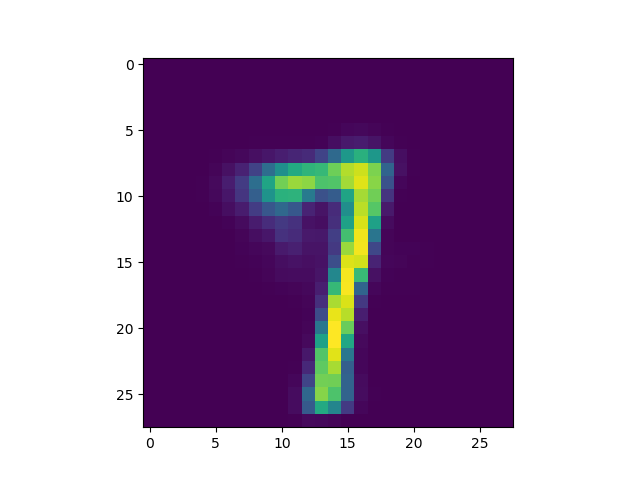} &
    \hspace{-1cm}\includegraphics[width=0.28\textwidth]{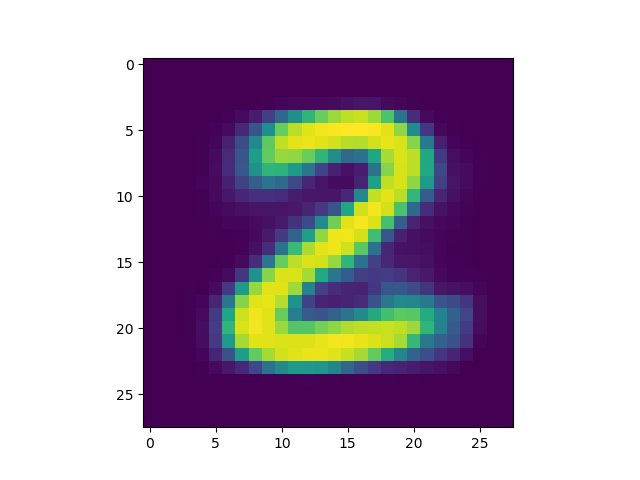} & \hspace{-1cm}\includegraphics[width=0.28\textwidth]{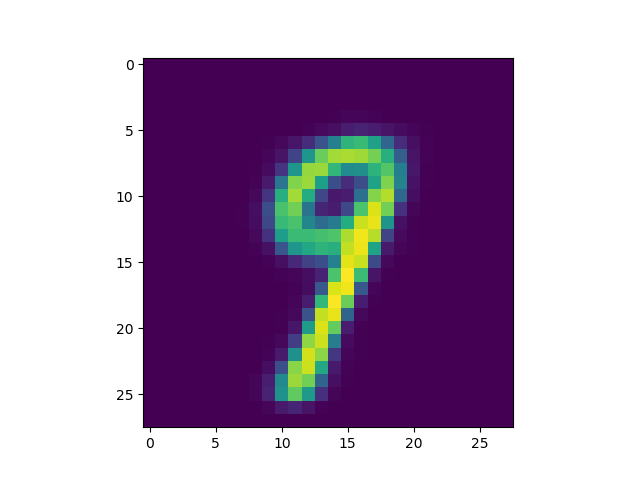} & \hspace{-1cm}\includegraphics[width=0.28\textwidth]{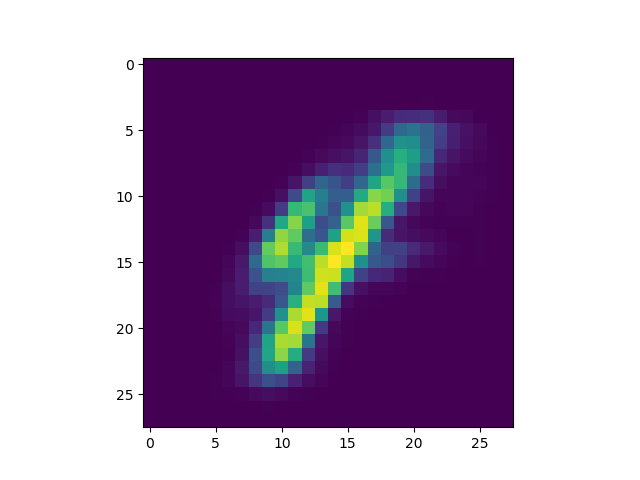} \\
    \end{tabular}
    \caption{Queries (first row) and answers (second row) of {NeuroDB} and true answers (third row)}
    \label{fig:mnist_queries}
    \end{minipage}

\end{figure*}

\begin{figure*}
    \centering
    \begin{tabular}{c c c c c c }
    \hspace{-0.5cm}\includegraphics[width=0.2\textwidth]{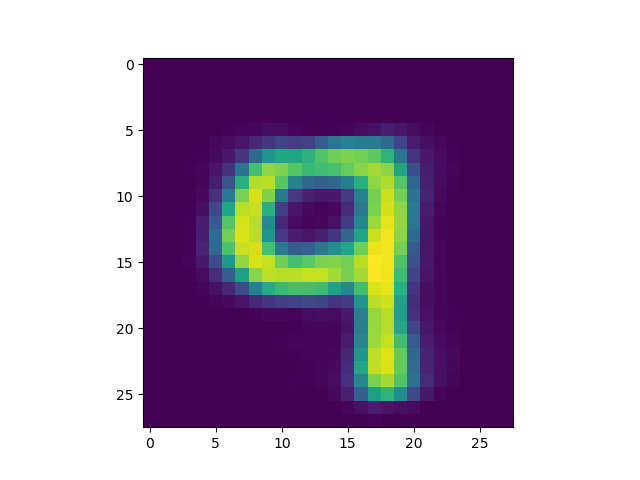} &
    \hspace{-1cm}\includegraphics[width=0.2\textwidth]{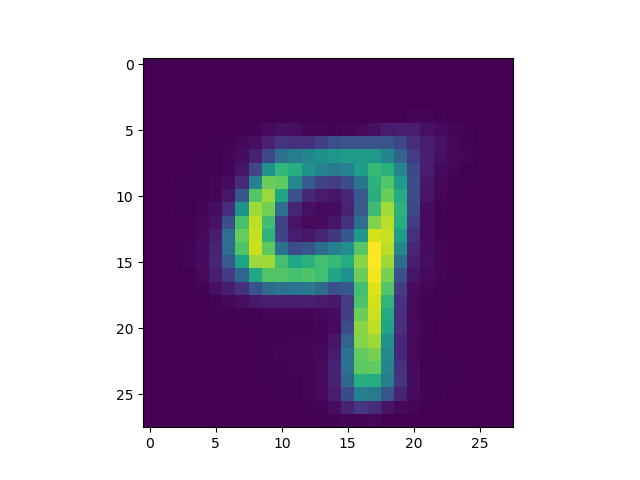} & \hspace{-1cm}\includegraphics[width=0.2\textwidth]{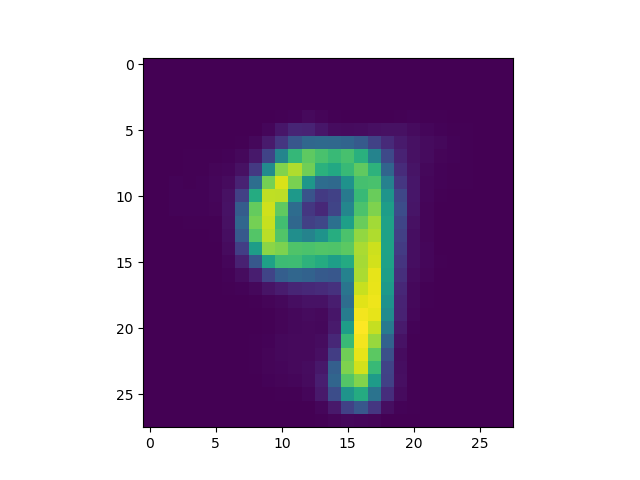} & \hspace{-1cm}\includegraphics[width=0.2\textwidth]{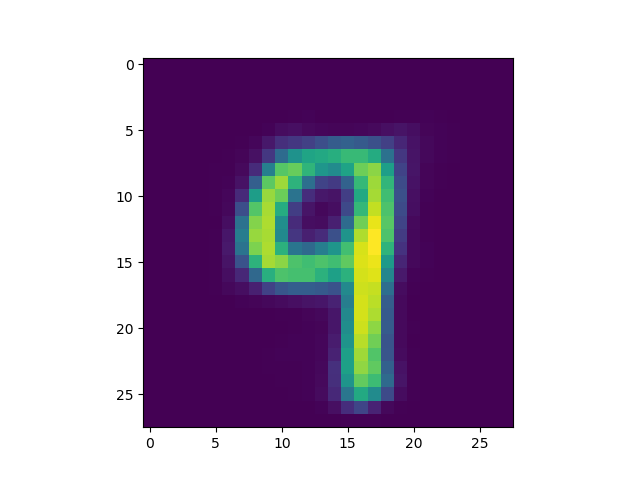} & \hspace{-1cm}\includegraphics[width=0.2\textwidth]{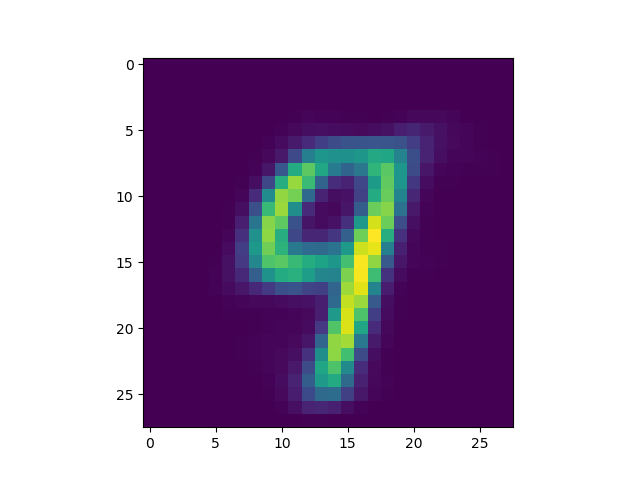} & \hspace{-1cm}\includegraphics[width=0.2\textwidth]{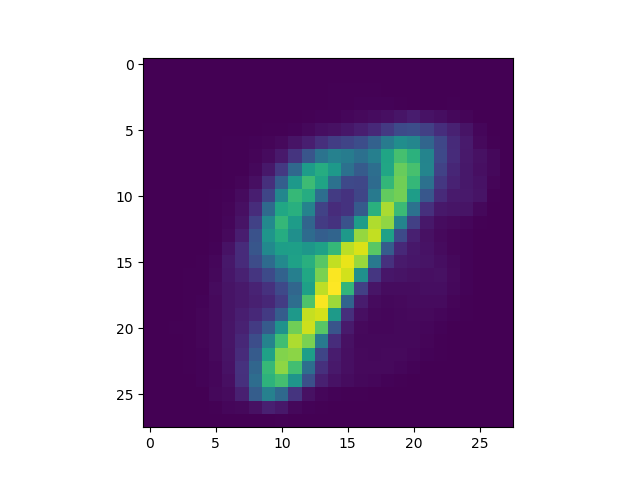}   \\
    \hspace{-0.5cm}\includegraphics[width=0.2\textwidth]{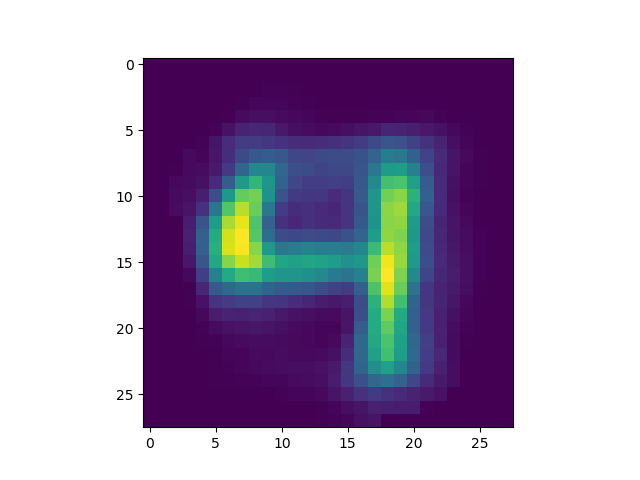} &
    \hspace{-1cm}\includegraphics[width=0.2\textwidth]{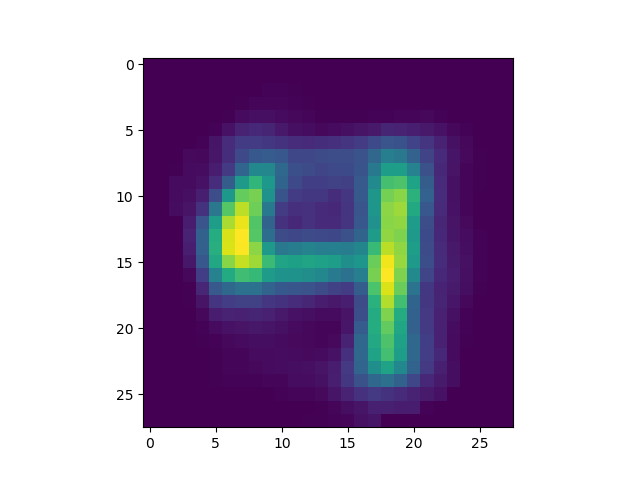} & \hspace{-1cm}\includegraphics[width=0.2\textwidth]{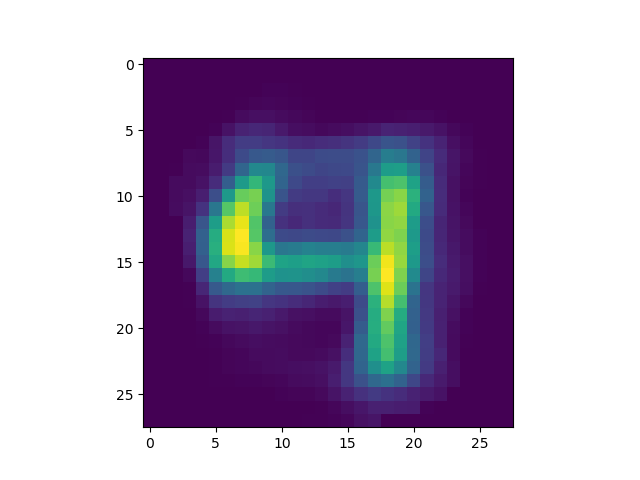} & \hspace{-1cm}\includegraphics[width=0.2\textwidth]{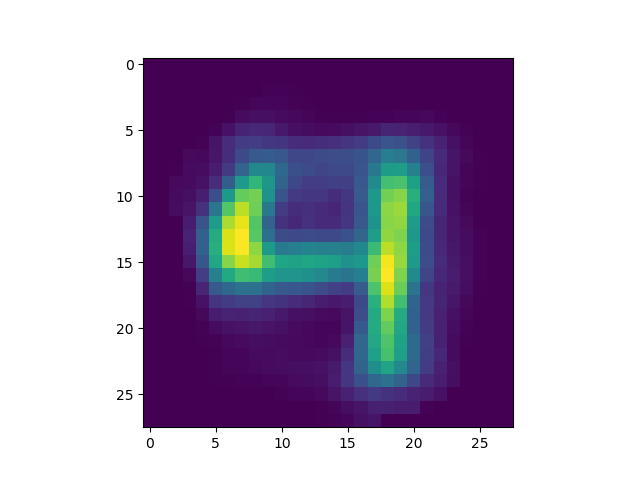} & \hspace{-1cm}\includegraphics[width=0.2\textwidth]{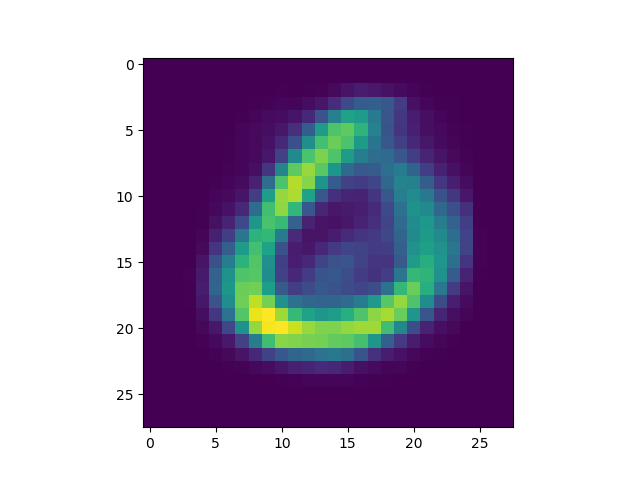} & \hspace{-1cm}\includegraphics[width=0.2\textwidth]{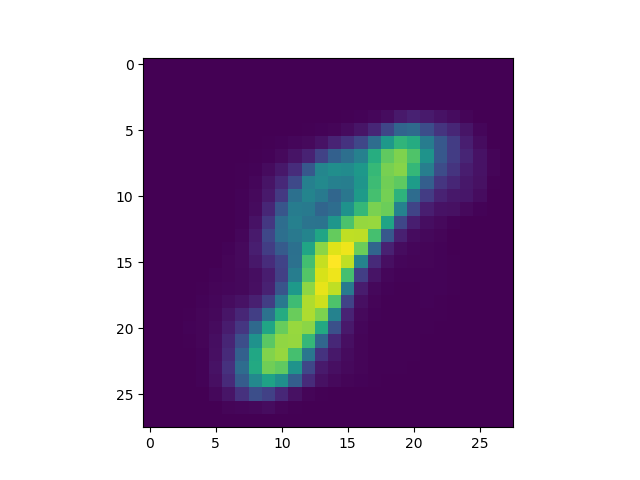} 
    \end{tabular}
    \caption{Queries (first row) and answers (second row) of {NeuroDB} trained on a dataset without the digit 9}
    \label{fig:nine_exp}
\end{figure*}

\subsection{Learned patterns for Nearest Neighbour Query}\label{appx:NN_exp}
We use a smaller data set to study how {NeuroDB} performs nearest neighbour queries to provide interesting insights into what NeuroDB learns.  

\noindent\textbf{Dataset}. We used the mnist dataset created by \cite{loosli-canu-bottou-2006} which contains 28$\times$28 gray-scale pixel hand-written digits (i.e. each image has 784 dimensions). We use a variational auto-encoder (VAE) \cite{kingma2013auto} to first learn a 30-dimensional representation of each image. Then, we create a databases, $D$, containing 10 different digits. We use the rest of the images in the mnist dataset to be our training and testing sets. 5 of the images (digits) in the database are shown in Fig.~\ref{fig:mnist_db} (note that the database contains 5 more images not shown).

\subsubsection{NeuroDB and Feature Learning}\hfill\\
\noindent\textbf{Goal}. In this experiment our goal is to (1) gain insight about the output of {NeuroDB} and (2) discuss the potential of {NeuroDB} in helping machine learning methods perform better feature learning. We emphasize that this experiment is not a simple application of VAEs, but rather shows the potential of {NeuroDB} in helping VAEs learn better features. The query can be thought of as a \textit{style transfer} task where the goal is to replace an image with the same digit in $D$. This can be done with a nearest neighbour query if a good representation of each image is learned. 

\noindent\textbf{Results and Discussion}.  Fig.~\ref{fig:mnist_queries} shows multiple input queries, their corresponding output of {NeuroDB} and the true nearest neighbour (the images are plotted as the output of the VAE). First, we observe that, the output of {NeuroDB} and the true results are visually indistinguishable when decoded, even though their representations aren't exactly the same. For instance, in the first row, the relative error (as defined in Sec. \ref{sec:exp:dist_NN}) is 0.026. This shows that small approximation error can be tolerated in practice. 

Second, we observe that in the fourth column, the digit 1 is mapped to the digit 4. This can be attributed to the fact that the VAE has not learned a good representation for the digit 1, and has mapped it to a location closer to the digit 4 in the feature space. We note that this is not caused by the error in {NeuroDB} as performing nearest neighbour search with no approximation error (shown in the third row) also returns digit 4 and not 1. That is, the nearest neighbour query shows the problem in the feature learning. 

Although {NeuroDB} isn't at fault for observing this issue in feature learning, it can be useful in fixing it. {NeuroDB} provides a differentiable nearest neighbour query operator, and thus can be backpropogated through (in contrast with combinatorial methods that perform nearest neighbour search). For instance, a loss on distance to nearest neighbour (which can be calculated with {NeuroDB}) can enforce the representations being similar to to the digits in the database. We leave the potential of using {NeuroDB} in feature learning for the future work, but we briefly mention that another simple potential use-case is to use {NeuroDB} as part of the encoder, that is, to consider the output of the {NeuroDB} as the final encoding. If good enough representations are learnt by the VAE, {NeuroDB} can help create a \textit{unique} representation for each digit, which can make a downstream classification task easier.  

\subsubsection{What is being learnt?}\hfill\\
\noindent\textbf{Goal}. In this section we address the question of whether the neural network is learning any interesting patterns between the input and its nearest neighbour. To do so, we perform a similar experiment as above, but remove the digit 9 from the dataset during training. That is, digit 9 is removed from the database as well as the training set. The rest of the training is done as before. At test time, we examine what the neural network outputs when digit 9 is being input. 

\noindent\textbf{Results and Discussion}. Fig.~\ref{fig:nine_exp} shows the results of this experiment. The first row shows the queries and the second row shows the output of the neural network. An interesting observations can be made from the results. The neural network is able to output a digit 9, when 9 is input to the model. Although this dose not always happen (e.g., in the last two columns of Fig.~\ref{fig:nine_exp}) the fact that it is possible is in itself significant. This is because a combinatorial method used to answer the query will always output another digit, i.e., a digit from 0-8 given the digit 9 as query (because 9 is not in the database, and the output of the combinatorial method is always in the database). In contrast, {NeuroDB} has learned a mapping for the nearest neighbour query, which is general enough so that a digit 9 as an input is still mapped to a digit 9. This behaviour is beneficial when there is missing data in the database. This also shows that {NeuroDB} is not merely memorizing training instances, but rather learning generalizable patterns. Another interesting observation is that the digit 9s that are output by the model are similar (e.g., the images depicted in the first 4 columns of Fig.~\ref{fig:nine_exp})

\end{document}